\documentclass{article}
\usepackage{arxiv}
\usepackage{hyperref}
\usepackage[utf8]{inputenc} 
\usepackage[T1]{fontenc}    
\usepackage{url}            
\usepackage{booktabs}       
\usepackage{amsfonts}       
\usepackage{nicefrac}       
\usepackage{microtype}      
\usepackage{amsmath}
\usepackage{hyperref}

\usepackage{bm}
\usepackage{graphicx}
\usepackage{sidecap}
\usepackage[toc,page]{appendix}
\usepackage{float}
\usepackage{amsthm}
\usepackage{bbm}
\usepackage{cancel}
\usepackage{multirow}
\usepackage{natbib}
\usepackage{xcolor}
\usepackage{cleveref}

\usepackage[normalem]{ulem}

\def\pilotdays{D_{0}}
\def\followupdays{D_{1}}
\def\Bern{\mathrm{Bernoulli}}
\def\betad{\mathrm{Beta}}

\def\NBP{\mathrm{NBP}}
\def\SSP{\mathrm{SSP}}
\def\NN{\mathbb{N}}

\def \unitidx {n}

\def \dayidx {d}
\def\EE{\mathbb{E}}

\def \genericunit {\omega_{\unitidx}}
\def \genericcount {A_{\dayidx, \unitidx}}
\newcommand{\obsnews}[2]{u_{#1}^{(#2)}}
\newcommand{\news}[2]{U_{#1}^{(#2)}}
\newcommand{\prednews}[2]{\hat{U}_{#1}^{(#2)}}
\newcommand \triggerdata[1]{Z_{#1}}
\newcommand \daytrigger{\triggerdata{\dayidx}}

\newcommand \pr{\mathrm{pr}}
\newcommand{\tail}{\sigma}
\newcommand{\unittotal}{N}
\newcommand{\NegBin}[2]{\mathrm{NegBin}\left(#1, #2\right)}
\newcommand{\tNegBin}[3]{\mathrm{tNegBin}\left(#1, #2;#3\right)}
\newcommand{\RE}{\mathbb{R}}
\newcommand{\ind}{\mathbbm{1}}
\newcommand{\freq}{j}
\newcommand{\de}{\mathrm{d}}
\newcommand{\randommeasure}{\Theta}
\newcommand{\numsuccess}{r}
\newcommand{\LP}{\mathrm{LP}}
\newcommand{\CRM}{\mathrm{CRM}}
\newcommand{\rate}{\theta}
\newcommand{\unitrate}{\rate_{\unitidx}}
\newcommand{\prior}{\mathcal{M}}
\newcommand{\likelihood}{\ell}
\newcommand{\tagg}{X}
\newcommand{\tilting}{c}
\newcommand{\mass}{\beta}
\newcommand{\SBSP}{\mathrm{SBSP}}

\newcommand{\iid}{\stackrel{\mbox{\scriptsize i.i.d.}}{\sim}}
\newtheorem{theorem}{Theorem}
\newtheorem{corollary}[theorem]{Corollary}

\newtheorem{remark}{Remark}
\newtheorem{proposition}[theorem]{Proposition}

\DeclareMathOperator*{\argmax}{arg\,max}
\DeclareMathOperator*{\argmin}{arg\,min}
\begin{document}

\title{\Large{Improved prediction of future user activity in online A/B testing}}

\author{\begin{tabular}{cccc} 
Lorenzo Masoero & 
Mario Beraha & 
Thomas Richardson & 
Stefano Favaro \\ 
 & 
\small{\textnormal{Dept. of Mathematics}} & 
\small{\textnormal{Dept. of Statistics}} &
\small{\textnormal{Dept. of Economics and Statistics}} \\ 
\small{\textnormal{Amazon.com}} & 
\small{\textnormal{Politecnico di Milano}} & 
\small{\textnormal{University of Washington}} &
\small{\textnormal{University of Torino}}\thanks{Also affiliated with {Collegio Carlo Alberto}}
\\ 
\small{\texttt{masoerl@amazon.com}} & \small{\texttt{mario.beraha@polimi.it}} & 
\small{\texttt{thomasr@uw.edu}} & 
\small{\texttt{stefano.favaro@unito.it}} 
\end{tabular}}
\newcommand{\arxiv}{1}
\maketitle

\begin{abstract}
    In online randomized experiments or A/B tests, accurate predictions of participant inclusion rates are of paramount importance. These predictions not only guide experimenters in optimizing the experiment's duration but also enhance the precision of treatment effect estimates. In this paper we present a novel, straightforward, and scalable Bayesian nonparametric approach for predicting the rate at which individuals will be exposed to interventions within the realm of online A/B testing. Our approach stands out by offering dual prediction capabilities—it forecasts both the quantity of new customers expected in future time windows and, unlike available alternative methods, the number of times they will be observed. 
    We derive closed-form expressions for the posterior distributions of the quantities needed to form predictions about future user activity, thereby bypassing the need for numerical algorithms such as Markov chain Monte Carlo.
    After a comprehensive exposition of our model, we test its performance on experiments on real and simulated data, where we show its superior performance with respect to existing alternatives in the literature.
\end{abstract}

\section{Introduction}

The problem of predicting the size of a population from which random samples are drawn has a long history in the statistics literature.
Originally motivated by applications in ecology, where the goal is typically to determine the number of distinct species of animals within a population \citep{fisher1943relation, good1953population, burnham1979robust}, a variation of this problem has recently received considerable attention also in the genomics literature, where scientists are interested in predicting the number of future rare variants to be observed within a genomic study \citep{ionita2009estimating, zou2016quantifying, chakraborty2019using,masoero2022more}. 

Inspired by this recent literature, we here consider the setting of online A/B testing (or online randomized controlled trials). A/B tests are widely used in the tech industry to measure the effectiveness of new interventions on key performance indicator of interest \citep{gupta2019top}. In practice, A/B tests are ``online'' randomized trials in which online users are randomly exposed to either the old ``control'' experience, or the new ``treatment'' experience. After running the trial for a given period of time, experimenters perform statistical analyses to determine whether to adopt the new intervention.

In this context, when planning the execution of the experiment, an estimate of the total activity that is going to be recorded throughout the experiment is an important piece of information. 
In particular, accurate prediction of the total number of users that will be active or will ``trigger'' in the experiment, together with the rate or frequency with which they will come back or ``re-trigger'', can help experimenters determine the optimal duration of the experiment, and potentially help them form improved estimates of the impact of the intervention. 
For example, on the basis of the data collected during a ``pilot'' period during the first days of the experiments, experimenters might expect to collect enough information for the task at hand in the subsequent week.
Conversely, experimenters might observe few, relatively inactive users during the pilot period. In this case, they might have to run the experiment longer. Indeed, experimenters  typically prefer to run an experiment for the shortest time possible, as experiments are expensive to maintain, and are responsible for a number of risks. For example, if the experiment is faulty, it might cause an inconsistent experience for users. Furthermore, even if the experiment is properly set up, running it for a long time is in general suboptimal: if the treatment induces a worse experience than control, users will be dissatisfied; on the other hand, if the treatment is effective, running an experiment for a long time will cause delay in the adoption of the beneficial intervention. 
On the other hand, running an experiment for a short time comes with the risk of collecting data that is less representative of the whole population that would be targeted by the intervention. Indeed, users who are more active will be over-represented in the sample first triggering in the experiment in the initial phases of the experiment compared to later periods. 

Experimenters then face a tradeoff: should they reduce experimental risks and costs, and run the experiment for a shorter amount of time, forming their estimates using a smaller subset of individuals, or should they run the experiment for a longer period of time, facing larger costs and risks but potentially reducing the risk of forming their estimates on the basis of a non-representative subpopulation?

Similar tradeoffs arise in classical experimental design (species discovery in capture-recapture problems \citep{good1953population, burnham1979robust}, genomics designs \citep{ionita2010optimal, masoero2022more, shen2022double}). These problems, however, are particularly pressing in online A/B testing, where experiments are run for a short time, and the outcome of an experiment directly informs critical decision-making. Motivated by this necessity, there has been a recently growing literature on sample size prediction and duration recommendation for online experiments \citep{richardson2022bayesian, wan2023experimentation}. 

In this paper we add to this literature by introducing a novel, simple and scalable Bayesian nonparametric approach which allows accurate prediction of the number of future users triggering in the experiment. In contrast to competing methods, our method also allows direct prediction of the future rates at which previously seen and yet-to-be seen users will trigger in the experiment. 

\section{A nonparametric Bayesian model to predict future activity in online A/B tests} \label{sec:model}

In this section, we provide details on our proposed  methodology for the problem of predicting future user activity in online A/B tests.
We consider the setting in which an experimenter has already run the experiment for an initial ``pilot'' period of $\pilotdays$ days, and is interested in forecasting future activity in a ``follow-up'' time window of $\followupdays$ days.
Our model builds on the recent contributions of \citet{camerlenghi2022scaled, beraha2023transform}, and more generally our setup closely follows the approach of \citet{masoero2022more}, which considered a similar prediction and experimental design problem in the context of rare-variants discovery in genomic studies. 

Formally, consider an experimenter who is running an A/B test to estimate the effectiveness of an intervention on an online platform. The experimenter might be trying to measure the impact on user engagement due to a new configuration (``treatment'') of the website with respect to the old default configuration (``control''). To do so, they run an online experiment or A/B test: that is, over a given time window, they assign a random ``tag'' $\tagg_\unitidx \in \{0,1\}$ to each user $\unitidx=1,2,\ldots$ first ``triggering'' in the experiment (e.g., by flipping an unbiased coin, $\tagg_\unitidx\sim \Bern(1/2)$). If $\tagg_\unitidx=0$, user $\unitidx$ will be exposed to the ``control'' experience (status quo) throughout the experiment, while if they are assigned $\tagg_\unitidx=1$, they will be exposed to the ``treatment'' experience throughout the experiment. While running the experiment, the experimenter collects ``activity data'' for the relevant metrics of interest (e.g., user engagement). After collecting triggering and activity data on the user engagement from both the control and treatment group, the experimenter performs a statistical analysis to assess the causal impact of the intervention on the population.
See \citet{gupta2019top, masoero2023leveraging} for recent overviews. 
\begin{figure}
    \centering
    \includegraphics[width = .5\textwidth]{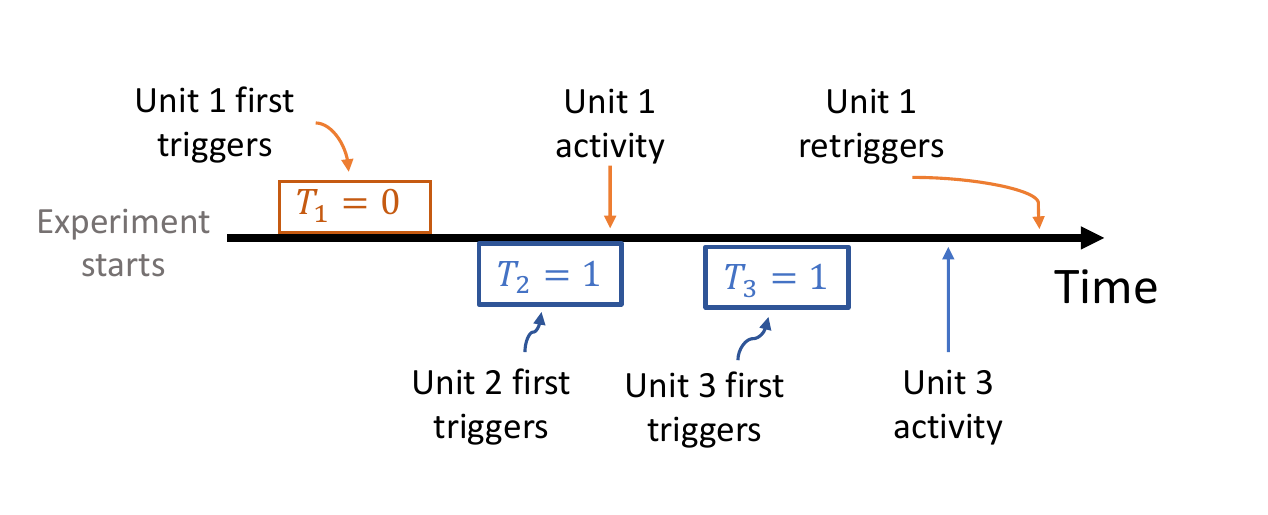}
    \caption{Sketch of an online A/B tests. Three units first trigger and subsequently re-trigger in the experiment. 
    }
    \label{fig:ab_sketch}
\end{figure}

We now assume that the online experiment has already run for the initial pilot duration of $\pilotdays$ days, during which the experimenter has collected re-trigger data.
Our proposed method, like existing alternatives in the literature, predicts future triggering activity in the control and treatment arm separately. Therefore, in what follows we tailor our discussion to a single treatment arm.  

Suppose that  we have observed a total of $0<\unittotal_{\pilotdays}<\infty$ users in the treatment arm of interest in the first $\pilotdays$ days. 
We denote with $\genericunit$ the unique identifier of the $\unitidx$-th unit observed in the experiment, in order of appearance (e.g., a pseudonymized id uniquely identifying a user), and with $\genericcount \in \{0,1,\ldots\}$ the count of the number of times that the $\unitidx$-th unit ``re-triggered'' in the experiment during the $\dayidx$-th day. 
Then, our observed data after $\pilotdays$ days consists of $\{(\omega_\unitidx, A_{\dayidx, \unitidx}), {\dayidx=1, \ldots, \pilotdays, \unitidx =1, \ldots, \unittotal_{\pilotdays}}\}$. We represent the data collected during day $\dayidx$ as a measure
\begin{equation}
    \triggerdata{\dayidx}(\cdot) = \sum_{\unitidx = 1}^{\unittotal_{\pilotdays}} \genericcount \delta_{\genericunit}(\cdot), \label{eq:data}
\end{equation}
which pairs each daily re-trigger count $\genericcount$ of the $\unitidx$-th unit on the $\dayidx$-the day together with the label $\genericunit$ of the $\unitidx$-th unit. Here $\delta$ is the Dirac measure, so that for a given pseudonmized id $\omega_{\unitidx}$, $\triggerdata{\dayidx}(\genericunit)$ equals the count $\genericcount$ of number of times unit $\unitidx$ triggered on day $\dayidx$.
Given days $\dayidx_0 < \dayidx_1$, we denote with $\triggerdata{\dayidx_0 : \dayidx_1}$ the collection of re-trigger data for all days between $\dayidx_0$ and $\dayidx_1$, $\triggerdata{\dayidx_0 : \dayidx_1}:=(\triggerdata{\dayidx_0}, \triggerdata{\dayidx_0+1},\ldots,\triggerdata{\dayidx_1})$. Given the observed data $\triggerdata{1: \pilotdays}$, we leverage a Bayesian approach to predict future triggering and re-triggering activity that will be observed in experimental days $d>\pilotdays$. In particular, given a latent parameter $\randommeasure$, we pose a generative model for the observed data $\triggerdata{1:\pilotdays}$. This consists of a likelihood function $\pr(\triggerdata{1: \pilotdays} \mid \randommeasure)$ for the trigger data, and an adequate prior distribution $\pr(\randommeasure)$ for the latent parameter.

One of the immediate challenges that needs to be addressed when positing the aforementioned Bayesian model is that the total number of users $\unittotal_{\pilotdays+\followupdays}$ that will be observed in the experiment is unknown. In practice, the triggering logic and experimental design induce a finite upper bound on the total number of units who can trigger in any given experiment. However, such upper bound might vary dramatically across experiments, and more generally might be hard to model or estimate. Therefore, in what follows, to circumvent the necessity of specifying an artificial upper bound, we imagine that there exists a countable infinity of units who can trigger in the experiment. In turn, this allows us to re-write our data in \Cref{eq:data} as the measure
\[
    \triggerdata{\dayidx} = \sum_{\unitidx = 1}^{\infty} \genericcount \delta_{\genericunit},
\]
where $\genericcount=0$ for all $\unitidx>\unittotal_{\pilotdays+\followupdays}$. Further, we assume that for every unit $\unitidx$ and every duration of the experiment $D$, the distribution of counts $(A_{1,\unitidx}, \ldots, A_{D,\unitidx})$ is exchangeable. That is, for any finite permutation $\pi:\{1,\ldots,D\}\to\{1,\ldots,D\}$, it holds that $\pr(A_{1,\unitidx}, \ldots, A_{D,\unitidx}) = \pr(A_{\pi(1), \unitidx}, \ldots, A_{\pi(D), \unitidx})$ for every unit $\unitidx$. Roughly, this amounts to assuming that the unit-specific probability of re-triggering is constant over time.
Though this assumption might be violated  in many experiments, we show that in practice, on real data, we are still able to form accurate estimates for our quantities of interest.
In turn, because the re-trigger counts $A_{1:D,\unitidx}$ are exchangeable, they can be seen as conditionally identically and independent distributed given a latent random ``rate'' $\unitrate$. We represent the unit-level rates $\unitrate$ coupled together with the unit-level identifiers $\genericunit$ in a measure 
\[
    \randommeasure = \sum_{\unitidx=1}^{\infty} \unitrate \delta_{\genericunit}.
\]
We adopt a Bayesian nonparametric approach, treating the infinite-dimensional parameter $\randommeasure$ as random (i.e., as a random measure), and endowing it with an adequate prior distribution $\mathcal{M}$. We write $\randommeasure\sim \mathcal{M}$. Conditionally on $\randommeasure$, we assume that the re-trigger counts are distributed according to a likelihood function $\ell$, where the count $\genericcount$ only depends on $\theta_\unitidx$:
\[
    \genericcount \mid \randommeasure \sim \ell( \cdot \mid\unitrate),
\]
independently across $\unitidx$ and independently and identically distributed (i.i.d.) across $\dayidx=1,2,\ldots$ for the same unit $\unitidx$. We denote the distribution of $Z_{\dayidx} \sim \LP(\ell, \randommeasure)$ --- and we refer to it as the ``likelihood process'' (LP) with likelihood $\ell$ and parameter $\randommeasure$ \citep{nguyen2023independent}. Our Bayesian nonparametetric hierarchical model can be summarized as:
\begin{equation}
        \Theta \sim \prior, \quad
        \daytrigger \mid \Theta \sim \LP(\ell, \Theta)\;\text{for}\; d \ge 1. \label{eq:model_general}
\end{equation}

\subsection{The negative-binomial stable-beta scaled-process}

In what follows, we tailor the model in \Cref{eq:model_general} by specifying a likelihood function $\likelihood$ for the likelihood process $\LP(\likelihood,\randommeasure)$ --- which describes the distribution responsible for generating counts $\genericcount$ --- and a prior distribution $\prior$ for $\randommeasure$ --- which is responsible for the distribution of the user rates $\unitrate$. We assume that conditionally on $\randommeasure$ and given a fixed scalar parameter $\numsuccess > 0$, it holds
\[
    \genericcount \mid \randommeasure \sim \NegBin{r}{\unitrate},
\]
independently across $\unitidx$ and i.i.d.\ across $\dayidx$. That is, 
\[
    \pr(\genericcount = x \mid \randommeasure) = \binom{x+r-1}{x} \unitrate^x (1-\unitrate)^ r \ind(x \in \NN).
\]
To denote this distribution, we write $\triggerdata{\dayidx} \mid \randommeasure \sim \NBP(\randommeasure, r)$.
We let the random measure $\randommeasure$ be distributed according to a stable-beta scaled-process ($\SBSP$), a prior first introduced in \citet{camerlenghi2022scaled}. We summarize its construction here, and refer the reader to \citet{camerlenghi2022scaled, beraha2023transform} for additional details. 
\begin{remark}[Model specification]
    Our choices of the $\SBSP$ prior distribution and the associated negative binomial likelihood model are informed by computational considerations. 
    Namely, as we show in this section, these choices allow us to derive exact posterior and predictive characterizations for the model (\Cref{sec:posterior}). 
    In turn, this enables us to perform inference at scale, incurring limited computational cost, while retaining the benefits of a rich and flexible nonparametric model.
\end{remark}
To obtain a SBSP, start by considering a homogeneous stable process: i.e., a discrete random measure of the kind $\randommeasure_0=\sum_{\unitidx \geq 1} \xi_\unitidx \delta_{\genericunit}$ on $\RE_+ \times \Omega$, where the collection $(\xi_\unitidx)_{\unitidx \geq 1}$ contains the jumps of a Poisson point process (PP) over $\RE_+$ characterized by the L{\'e}vy intensity measure 
\begin{equation}
    \nu(\de s) = \rho(s) \de s = \tail s^{-1-\tail} \ind(s>0) \de s, \label{eq:levy}
\end{equation} 
for $\tail \in (0, 1)$, and $\genericunit \iid P_0$ for $P_0$ a diffuse probability measure on a Polish space $\Omega$. In our development, the $\genericunit$ are immaterial and serve purely as unique tags for individual experimental units, so we henceforth let $P_0$ be the uniform measure on $[0,1]$.
The SBSP prior we consider is obtained by a suitable normalization of the jumps $(\xi_\unitidx)_\unitidx$. Denoting by $\Delta_0>\Delta_1>\ldots$ the sorted sequence of jumps $(\xi_\unitidx)_{\unitidx \ge 1}$, let $F_{\Delta_0}$ be the distribution of the largest jump $\Delta_0$, and  $f_{\Delta_0}$ the corresponding density. For the stable process of \eqref{eq:levy} this distribution has density \citep{ferguson1972representation},
\[
    f_{\Delta_0}(s) = \tail s^{-1-\tail}e^{-s^{-\tail}}\ind(s>0).
\]
Let now $\Delta_{0,h}$ be a random variable with density given by $f_{\Delta_{0,h}}(s) = h_{\tilting, \mass}(s)f_{\Delta_{0}}(s)$, where for $\tilting>0$ and $\mass > 0$,
\[
    h_{\tilting, \mass}(s) = \frac{\mass^{\tilting+1}}{\Gamma(\tilting+1)} s^{-\tilting \tail} \exp\left\{ - (\mass-1) s^{-\tail} \right\},
\]
with $\Gamma(\cdot)$ denoting the gamma function. 

Denote by $G_{t}$  the 
joint law of the infinite sequence $({\Delta_{\unitidx}}/{\Delta_0})_{\unitidx\ge 1}$ conditionally on $\Delta_0=t$. The scaled measure
\[
    \randommeasure = \sum_{\unitidx\ge 1} \frac{\Delta_{\unitidx}}{\Delta_0} \delta_{\genericunit} = \sum_{\unitidx \ge 1} \unitrate \delta_{\genericunit},
\]
is a SBSP prior if the weights $(\unitrate)_{\unitidx\ge 1}$ have distribution $G_{\Delta_{0,h}}$ i.e., their joint law is $\int G_t(\cdot) h_{\tilting, \mass}(t) f_{\Delta_{0}}(t) \de t$. 

We write $\randommeasure ~\sim~ \SBSP(\mass,\tail,\tilting)$. To sum up, the model we consider is
\begin{align} \label{eq:model}
\begin{split}
	\randommeasure &\sim \SBSP(\mass, \tail, \tilting) \\
        \daytrigger \mid \randommeasure  &\iid \NBP(\randommeasure, r), \quad \dayidx \ge 1. 
\end{split}
\end{align}

\subsection{Posterior and predictive characterization of the negative binomial scaled-stable process} \label{sec:posterior}

In this section, we characterize the predictive structure of the Bayesian nonparametric model of \Cref{eq:model}. In turn, we show in \Cref{sec:theory_prediction} how by leveraging this posterior predictive structure we can derive practical, simple and scalable Bayesian nonparametric  predictors for future re-triggering rates of new and old users in the experiment. Thanks to these exact posterior representations, we will be able to avoid the need of using more expensive inference schemes (e.g., Markov Chain Monte Carlo). See \Cref{sec:proofs} for proofs of our results.
We first provide the posterior distribution of $\Delta_{0, h}$. Then, we give the expressions for the posterior distribution of $\Theta$ and the predictive distribution of the model conditional on $\Delta_{0, h}$, for which simple expressions are found.
In \Cref{sec:theory_prediction} we show how integrating the predictive distribution against the posterior of $\Delta_{0, h}$ leads to tractable estimators for future users' activity.

\begin{proposition}[Distribution of the largest jump]\label{prop:jump}
    Let $f:=f_{\Delta_{0, h}\mid \triggerdata{1:\pilotdays}}$ denote the posterior density of the largest jump. Under the model of \Cref{eq:model} and given $\triggerdata{1:\pilotdays}$, it holds
\begin{equation*}
    f(\eta) \propto 
    e^{\left\{ - \eta^{-\tail} \left[\mass + \psi_0^{(\pilotdays)}\right] \right\}} \eta^{-\tail(\unittotal_{\pilotdays}+\tilting+1)-1}
    \ind(\eta\in\RE_+).
\end{equation*}
That is, $\Delta_{0, h}^{-\tail}\mid \triggerdata{1:\pilotdays} \sim \mathrm{Gamma}(\unittotal_{\pilotdays}+\tilting+1, \mass + \psi_0^{(\pilotdays)})$. Here, letting $B(a,b)$ be the beta function with parameters $a$, $b$, we use the notation 
\[
    \psi_{x}^{(y)}= \tail\left[B(rx+1,-\tail) - B(r(x+y)+1,-\tail)\right].
\]    
\end{proposition}

\begin{proposition}[Posterior representation]\label{prop:posterior}
For the Bayesian model in \Cref{eq:model}, given $\triggerdata{1:\pilotdays}$, 
 $\Delta_{0, h}$, the posterior distribution of the random measure $\randommeasure$ is:
\begin{equation}
    \randommeasure \mid \Delta_{0, h}, \triggerdata{1:\pilotdays} \overset{d}{=} \tilde{\randommeasure} + \sum_{n=1}^{\unittotal_{\pilotdays}} J_\unitidx \delta_{W_\unitidx^\star}(\cdot). \label{eq:posterior}
\end{equation}
$\tilde{\randommeasure} \mid \Delta_{0, h}$ is a completely random measure (CRM, \citep{kingman1992poisson}) over 
$\Omega$ with almost surely distinct atoms,
    \[
        \tilde{\randommeasure} \mid \Delta_{0, h} \sim \CRM(\tilde{\nu}(\de s) \times P_0(\de \omega))
    \]
    with
    \[
        \tilde{\nu}(\de s) = (1-s)^{r\pilotdays}\Delta_{0, h}\rho (\Delta_{0, h}s)  \ind(s\in(0,1))\de s,
    \]
    with $\rho$ as in \Cref{eq:levy}.
    The other component is a discrete random measure which pairs fixed locations $W_{1}^\star,\ldots,W_{\unittotal_{\pilotdays}}^\star \in \Omega$ with random jumps $J_{1:\unittotal_{\pilotdays}}$. These jumps are independent and distributed according to:
    \begin{equation}
            J_{\unitidx} \mid \Delta_{0, h}, \triggerdata{1:\pilotdays} \sim \betad(m_{\pilotdays, \unitidx}-\tail, r{\pilotdays}+1), \label{eq:posterior_jumps}
    \end{equation}
    with $m_{\pilotdays, \unitidx} = \sum_{\dayidx=1}^{\pilotdays} \triggerdata{\dayidx, \unitidx}$ the total retrigger rate for unit $\unitidx$ in the first $\pilotdays$ days of the experiment.
\end{proposition}
We now characterize the predictive distribution of the model.
\begin{proposition}[Predictive representation]\label{prop:predictive}
    Consider the model in \Cref{eq:model}. Given $\triggerdata{1:\pilotdays}$ and $\Delta_{0, h}$, the distribution of $\triggerdata{\pilotdays+1}$ is given by: 
\begin{equation}
   \triggerdata{\pilotdays+1} \mid \triggerdata{1:\pilotdays}, \Delta_{0, h} \overset{d}{=} 
   \triggerdata{\pilotdays+1}' + 
   \sum_{n=1}^{\unittotal_{\pilotdays}} A_{\pilotdays+1, \unitidx} \delta_{W_\unitidx^\star}, \label{eq:predictive_characterization}    
\end{equation}
where, given $\tilde{\randommeasure}$ as in \Cref{prop:posterior},
 \[
    \triggerdata{\pilotdays+1}' \mid \tilde{\randommeasure}, \Delta_{0, h}  = \sum_{\unitidx\ge 1} A'_{\pilotdays+1,\unitidx}\delta_{W'_\unitidx} \sim \NBP(\tilde{\randommeasure}, r),
\]
and the random variables $A_{\pilotdays+1,\unitidx} \overset{\text{ind}}{\sim} \NegBin{r}{J_\unitidx}$ for $\unitidx = 1,\ldots,\unittotal_{\pilotdays}$, with $J_{\unitidx}$ distributed as in \Cref{eq:posterior_jumps}.
\end{proposition}

Finally, we also provide the marginal distribution of the observed sample.
\begin{proposition}[Marginal distribution of the observed sample] \label{prop:marg}
The marginal distribution under the model for $\pilotdays$ observations $\triggerdata{1:\pilotdays}$ is
\begin{align}
    \pr(\triggerdata{1:\pilotdays}) 
    &= \frac{\tail^{\unittotal_{\pilotdays}}\mass^{\tilting+1}}{\left(\mass + \psi_0^{(\pilotdays)} \right)^{\unittotal_{\pilotdays}+\tilting+1}} \frac{\Gamma(\unittotal_{\pilotdays}+\tilting+1)}{\Gamma(\tilting)} \nonumber\\
    &\quad \prod_{\unitidx=1}^{\unittotal_{\pilotdays}} \xi(A_{1:\pilotdays, \unitidx}, r, \tail), \label{eq:likelihood}
\end{align}
with
\begin{align*}
    \xi(A_{1:\pilotdays, \unitidx}, r, \tail)&:=\prod_{\dayidx:A_{\dayidx, \unitidx}>0}^D \left[ \binom{A_{\dayidx, \unitidx} + r - 1}{A_{\dayidx, \unitidx}}\right] \\
    &\quad B(\pilotdays r+1, m_{\pilotdays,\unitidx}-\tail+1).
\end{align*}
\end{proposition}
\subsection{Predicting future user re-trigger rates with the negative binomial scaled stable process} \label{sec:theory_prediction}

We now use the posterior characterization obtained in \Cref{sec:posterior} to derive our Bayesian nonparametric estimators of interest. First, we consider the total number of new (yet to be seen) users who are going to trigger in the experiment in $\followupdays$ future days given data $\triggerdata{1:\pilotdays}$ from $\pilotdays$ days:
\[
	\news{\pilotdays}{\followupdays}:= \sum_{\unitidx \ge 1} \ind\left( \sum_{\dayidx = 1 }^{\pilotdays} A_{\dayidx, \unitidx} = 0 \right)
	 \ind\left( \sum_{\dayidx = 1 }^{\followupdays}  A_{ \pilotdays+\dayidx, \unitidx} > 0  \right).
\]
\begin{proposition}[Number of new users] \label{prop:new_users}
	Under the model of \Cref{eq:model}, given trigger data $\triggerdata{1:\pilotdays}$ it holds
	\[
		\news{\pilotdays}{\followupdays} \mid \triggerdata{1:\pilotdays} \sim \NegBin{N_{\pilotdays} + \tilting + 1}{p_{\pilotdays}^{(\followupdays)}},
	\]
	with $p_{\pilotdays}^{(\followupdays)}:= \psi_{\pilotdays}^{(\followupdays)} / (\mass + \psi_{0}^{(\pilotdays + \followupdays)})$.
    We use the posterior expectation of $\news{\pilotdays}{\followupdays}$ as the Bayesian nonparametric predictor for the number of new users to be observed in $\followupdays$ additional days, 
    \begin{equation}
        \prednews{\pilotdays}{\followupdays} = \left(N_{\pilotdays} + \tilting + 1\right) \frac{p_{\pilotdays}^{(\followupdays)}}{1-p_{\pilotdays}^{(\followupdays)}}. \label{eq:BNP_pred}
    \end{equation}
\end{proposition}
Next, we derive the posterior distribution of $\news{\pilotdays}{\followupdays, \freq}$, the number of users who have not triggered before day $\pilotdays$ and will trigger exactly $\freq\ge 1$ times between $\pilotdays$ and $\followupdays$,  
\[
	\news{\pilotdays}{\followupdays, \freq}:= \sum_{\unitidx \ge 1} \ind\left( \sum_{\dayidx = 1 }^{\pilotdays} A_{\dayidx, \unitidx} = 0 \right)
	 \ind\left( \sum_{\dayidx = 1 }^{\followupdays}  A_{\pilotdays+\dayidx, \unitidx} = \freq  \right).
\]
This quantity is useful, as it provides experimenters with an estimate of the total activity that they can expect to see in future experimental days.
\begin{proposition}[Number of new users with frequency] \label{prop:new_retrigger}
	Under the model of \Cref{eq:model}, given trigger data $\triggerdata{1:\pilotdays}$,
	\[
		\news{\pilotdays}{\followupdays, \freq} \mid \triggerdata{1:\pilotdays} \sim \NegBin{N_{\pilotdays} + \tilting + 1}{p_{\pilotdays}^{(\followupdays, \freq)}},
	\]
	with 
        \[
            p_{\pilotdays}^{(\followupdays, \freq)}:= \rho_{\pilotdays}^{(\followupdays, \freq)} / (\mass + \psi_{0}^{(\pilotdays)} + \rho_{\pilotdays}^{(\followupdays, \freq)}),
            \]
            and 
            \[
                \rho_{\pilotdays}^{(\followupdays, \freq)}:=\binom{\freq+r\followupdays +1}{\freq}  \tail B\left[r(\pilotdays+\followupdays) +1, \freq - \tail \right].
            \]
\end{proposition}
Similarly, we provide the posterior distribution of future re-trigger rates of users  already observed in the pilot data.
\begin{corollary}[Future re-trigger rates, old users] \label{cor:old_retrigger}
Define 
\[
	S_{\pilotdays}^{(\followupdays)} := \sum_{\dayidx=1}^{\followupdays} \sum_{\unitidx=1}^{\unittotal_{\pilotdays}} A_{\pilotdays+\dayidx, \unitidx}.
\]
Under the model of \Cref{eq:model}, given $\triggerdata{1:\pilotdays}$
\[
	S_{\pilotdays}^{(\followupdays)} \mid \triggerdata{1:\pilotdays} \overset{d}{=}  \sum_{\unitidx=1}^{\unittotal_{\pilotdays}} \NegBin{r \followupdays}{J_\unitidx},
\]
with $J_{\unitidx} \overset{ind}{\sim} \mathrm{Beta}(m_{\pilotdays, \unitidx} - \tail, r \pilotdays + 1)$, as in \Cref{eq:posterior_jumps}.
\end{corollary}

\begin{corollary}[Total future re-trigger rates] \label{cor:total_retrigger}
Define the total future retrigger rate observed between day $\pilotdays$ and $\pilotdays+\followupdays$ 
\[
    T_{\pilotdays}^{(\followupdays)}:=\sum_{\unitidx \ge 0} \sum_{\dayidx=\pilotdays+1}^{\pilotdays+\followupdays} A_{\dayidx, \unitidx}.
\]
Under the model of \Cref{eq:model}, given $\triggerdata{1:\pilotdays}$, by combining \Cref{prop:new_retrigger} and \Cref{cor:old_retrigger}, it holds that
\[
    T_{\pilotdays}^{(\followupdays)} \mid \triggerdata{1:\pilotdays} \overset{d}{=} \sum_{\freq \ge 1} \freq \news{\pilotdays}{\followupdays, \freq} + S_{\pilotdays}^{(\followupdays)},
\]
where the distribution of $\news{\pilotdays}{\followupdays, \freq}$ was given in \Cref{prop:new_retrigger}
 and that of $S_{\pilotdays}^{(\followupdays)}$ was given in \Cref{cor:old_retrigger}.
\end{corollary}

\subsection{Marginal representation and sampling scheme} \label{sec:theory_sampling}

By leveraging the posterior and predictive representations presented in \Cref{sec:posterior,sec:theory_prediction} we can provide a practical algorithm to sample exactly observations $\triggerdata{1:\pilotdays}$ from the hierarchical model in \Cref{eq:model} via an ``urn-type'' scheme. Sampling the Observations $Z_{1:\pilotdays}$ doesn’t seem like a primary task for running an A/B test, but it is helpful to run simulation studies and assess the prediction abilities of our model. Let $\unittotal_{0}:=0$. Assume to have already sampled $\dayidx$ days of the experiment, $\dayidx = 0,\ldots,\pilotdays-1$. To generate $\triggerdata{\dayidx+1}$ given trigger data from the previous days, do
\begin{enumerate}
    \item Sample $\news{\dayidx}{1} \mid Z_{1:\dayidx} \sim \NegBin{\unittotal_{\dayidx}+\tilting+1}{p_{\dayidx}^{(1)}}$ new users, as per \Cref{prop:new_users}.
    \item For each new user $\unitidx = \unittotal_{\dayidx} + 1,\ldots, \unittotal_{\dayidx} + \news{\dayidx}{1}$, sample the corresponding re-trigger count $A_{\dayidx+1, \unitidx}$ i.i.d.\ from the probability mass functions
    \[
        \pr(A = \ell) \propto\frac{\Gamma(\ell+r-1)}{\Gamma(\ell)} B(\ell-\tail, r\dayidx+1) \ind_{\{1,2,\ldots\}}(\ell).
    \]
    \item For users $\unitidx=1,\ldots,\unittotal_{\dayidx}$ who already triggered in earlier days, sample counts from the marginal distribution
    \begin{equation*}
        \pr(A = \ell)\propto \binom{\ell + r - 1}{\ell} B(\ell+m_{\dayidx,\unitidx}-\tail, r\dayidx+1). 
    \end{equation*}
\end{enumerate}

    \subsection{Inference} \label{sec:inference}

We use the Bayesian nonparametric posteriors derived in \Cref{sec:theory_prediction} to directly form estimators by employing the posterior predictive mean of the statistic of interest (e.g., \Cref{eq:BNP_pred}). A challenge presented by these posteriors, however, is that they all depend on the hyperparameters $\mass, \tail, \tilting, r$ of the model (\Cref{eq:model}). In order to make practical use of these posteriors, then, we need to devise a procedure to use the data and \emph{fit} the hyperparameters. We here adopt an empirical Bayes approach. 

\paragraph{Maximum marginal likelihood} A natural option is to maximize the marginal likelihood $\pr(\triggerdata{1:\pilotdays})$ provided in \Cref{sec:posterior} with respect to the hyperparameters, i.e.\
\[
    \hat{\mass}, \hat{\tail}, \hat{\tilting}, \hat{r} = \argmax_{\mass, \tail, \tilting, r} \pr(\triggerdata{1:\pilotdays}; \mass, \tail, \tilting, r).
\]
This can be performed e.g.\ by employing numerical optimization techniques. In our codebase, we adopt the \texttt{scipy} implementation of the differential evolution algorithm \citep{storn1997differential}.
On simulated data drawn from the model, this approach correctly recovers the true underlying parameters (see, e.g., \Cref{fig:app_synthetic_log_like} in \Cref{sec:app_exp}).
\paragraph{Regression-based approach} Another option is to consider a regression problem to learn the hyperparameters. 
For example, assume that we want to predict the number of future users that will first trigger in the next $\followupdays$ days, using the point estimator of \Cref{eq:BNP_pred}. We fit the hyperparameters by solving the following regression problem:
\begin{equation*}
    \hat{\mass}, \hat{\tail}, \hat{\tilting}, \hat{r} = \argmin_{\mass, \tail, \tilting, r} \sum_{\dayidx = 1}^{\pilotdays - \dayidx_0} \left\{ \prednews{\dayidx_0}{\dayidx} - \obsnews{\dayidx_0}{\dayidx} 
\right\}^2,
\end{equation*}
where $\obsnews{\dayidx_0}{\dayidx}$ denotes the true (observed) number of new distinct users observed between day $\dayidx_0$ and day $\dayidx_0+\dayidx$, given $\triggerdata{1:\pilotdays}$. Here, $1\le \dayidx_0 < \pilotdays$, and $\dayidx_0 + \dayidx \le \pilotdays$.
In our experiments, we find that the choice of $\dayidx_0 = 1$ works well, although in different applications different choices might work better in practice (e.g., \citet[Section 4]{masoero2022more} recommends $\dayidx_0=\lfloor 2/3 \times \pilotdays \rfloor$).
Observe that $\prednews{\dayidx_0}{\dayidx}$ does not depend on the whole sample but only on the simple sufficient statistics, $\pilotdays, \unittotal_{\pilotdays}$ and the hyperparameters $\mass, \tail, \tilting, r$. 
Hence, an advantage of the regression approach over the maximum likelihood approach is that it is feasible even in the case in which only first trigger counts, and not re-trigger counts, are available. This scenario might occur on proprietary or large datasets, in cases in which, e.g.\ for privacy reasons or for reasons of scale, only aggregate summary statistics can be stored (e.g., the ASOS dataset later analysed \citet{liu2021datasets}).
\section{Experimental Results}

\subsection{Competing methods and evaluation framework}
We now move to the empirical evaluation of our proposed method. 
To benchmark its performance, we consider a number of alternative existing methods for sample size prediction: 
\begin{itemize}
    \item (J) Jackknife estimators up to the fourth order. These estimators first appeared in the capture-recapture literature in ecology \citep{burnham1979robust}. More recently, they have been employed in the genomics literature for future genomic variants prediction \citep{gravel2014predicting}.
    \item (LP) Nonparametric linear programming approach, developed by \citet{zou2016quantifying} in the context of rare variants discovery. Earlier versions of this algorithm have been employed to estimate the size of unknown populations, or the number of words contained in unknown vocabularies \citep{efron1976estimating}.
    \item (GT) A Good-Toulmin estimator. GT estimators have a long history in the statistics literature, dating back to the seminal work of \citet{good1953population, good1956number} in the context of rare species discovery. Here, we use a recent adaptation proposed by \citet{chakraborty2019using}.
    \item (BB, BG) the Bayesian method of \citet{ionita2009estimating}, developed in the context of rare genomic variants discovery, and the Bayesian model of \citet{richardson2022bayesian}, developed for the context of online A/B testing.
    \item (IBP, SSP) the Bayesian nonparametric methods of \citet{masoero2022more,camerlenghi2022scaled}, developed for prediction and optimal experimental design of large-scale genomic studies; also used in the context of power maximization of rare variants association studies \citep{masoero2021bayesian}.
\end{itemize}
We provide additional details on these competing methods and their implementation in \Cref{sec:app_competing}. 
Our experiments are performed in python on a MacBook M1 Pro, with 32 GB of RAM. To fit our method, we use the \texttt{scipy} implementation of the differential evolution algorithm \citep{2020SciPy-NMeth,storn1997differential}.\footnote{See \href{https://docs.scipy.org/doc/scipy/reference/generated/scipy.optimize.differential_evolution.html}{official documentation here}.}. In order to compare the performance of different methods, we use the following accuracy metric, already defined in \citet{camerlenghi2022scaled}:
\begin{equation}
    v_{\pilotdays}^{(\followupdays)} := 1 - \min\left\{ \frac{|\obsnews{\pilotdays}{\followupdays}-\prednews{\pilotdays}{\followupdays}|}{\obsnews{\pilotdays}{\followupdays}}, 1\right\} \in [0,1].
\end{equation}
This metric is equal to $1$ when the prediction is perfect and degrades to $0$ as its quality worsens.
$\obsnews{\pilotdays}{\followupdays}$ is the observed value of new users who first trigger between day $\pilotdays$ and day $\pilotdays+\followupdays$.  $\prednews{\pilotdays}{\followupdays}$ is a corresponding prediction stemming from one of the aforementioned methods, formed using trigger data from the first $\pilotdays$ days.
For our model, we use trigger data $\triggerdata{1:\pilotdays}$ to fit the hyperparameters following the empirical Bayes approach of \Cref{sec:inference}.  $\prednews{\pilotdays}{\followupdays}$ is the resulting posterior predictive mean of the random variable $\news{\pilotdays}{\followupdays}(\hat{\mass},\hat{\tail},\hat{\tilting},\hat{r})$ (as per \Cref{eq:BNP_pred}), where we  emphasize its dependence on the fitted hyperparameters. Similarly, we define a notion of accuracy for the prediction of the total retrigger rates $\hat{T}_{\pilotdays}^{(\followupdays)}$:
\[
    \tilde{v}_{\pilotdays}^{(\followupdays)} ~:=~ 1 - \min\left\{ \frac{|t_{\pilotdays}^{(\followupdays)}-\hat{T}_{\pilotdays}^{(\followupdays)}|}{t_{\pilotdays}^{(\followupdays)}}, 1\right\} \in [0,1].
\]
$t_{\pilotdays}^{(\followupdays)}$ is the observed total retrigger rate between $\pilotdays$ and $\pilotdays+\followupdays$
Unlike for the total number of new users, to the best of our knowledge no competing predictor exists for this quantity.

\subsection{Synthetic data}

\paragraph{Data from the model}

\begin{figure}[ht]
    \centering
    \includegraphics[width=\linewidth]{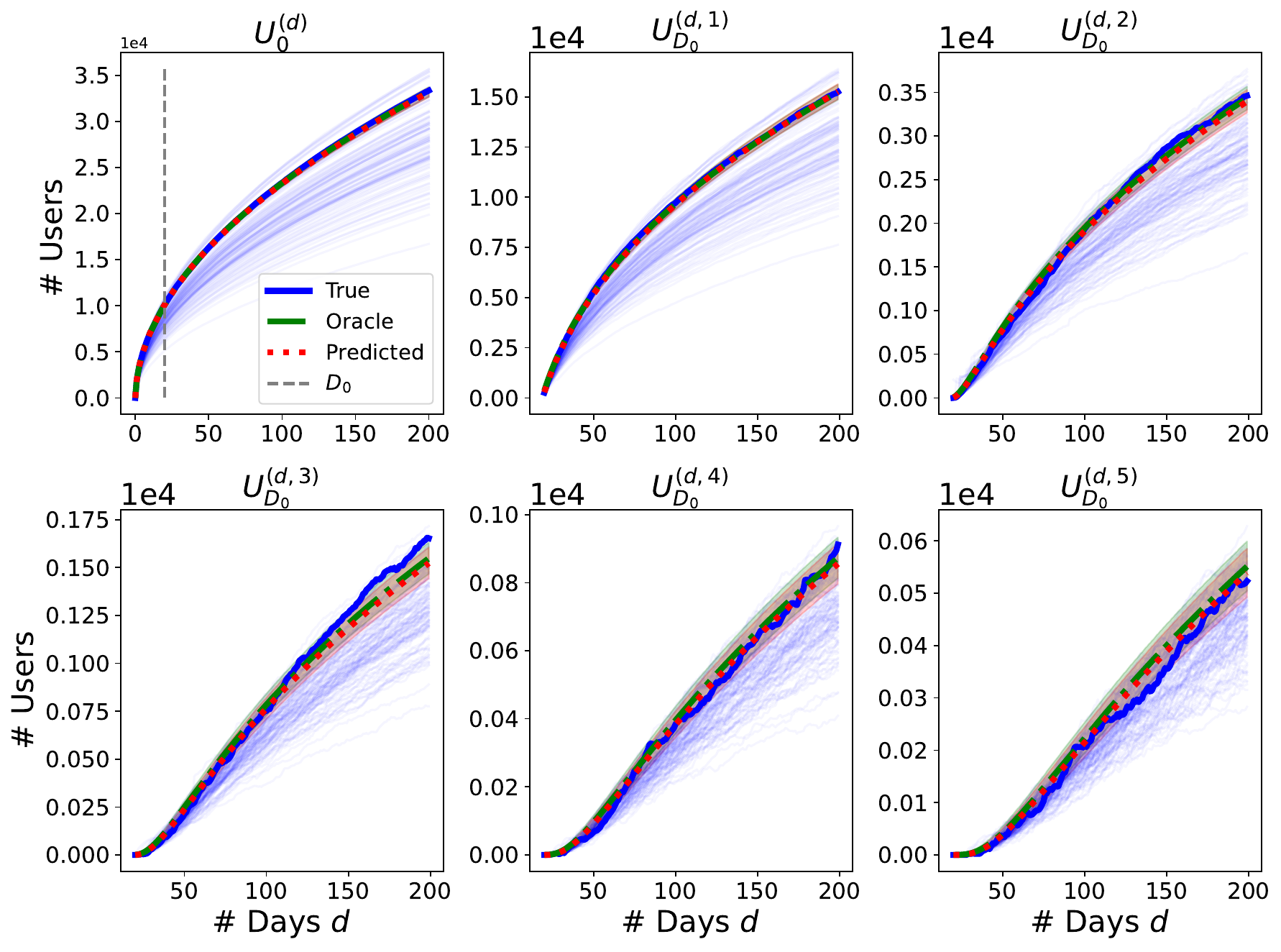}
    \caption{Prediction of the arrival of new users (top left, $\news{0}{\dayidx}$, other subplots, $\news{\pilotdays}{\dayidx, \freq}$ for $\freq = 1, \ldots, 5$). We plot the value of the statistic (vertical axis) as the number of days $\dayidx$ increases (horizontal axis). We compare the true value in the unobservable sample (solid blue line) to the predicted posterior mean with a centred 95\% credible interval using the fitted hyperparameters (red, dotted) and using the true hyperparameters (green, dash-dot). Shaded blue lines report additional $M=100$ draws from the model.}
    \label{fig:synthetic_1}
\end{figure}

\begin{figure}[ht]
    \centering
    \includegraphics[width=\linewidth]{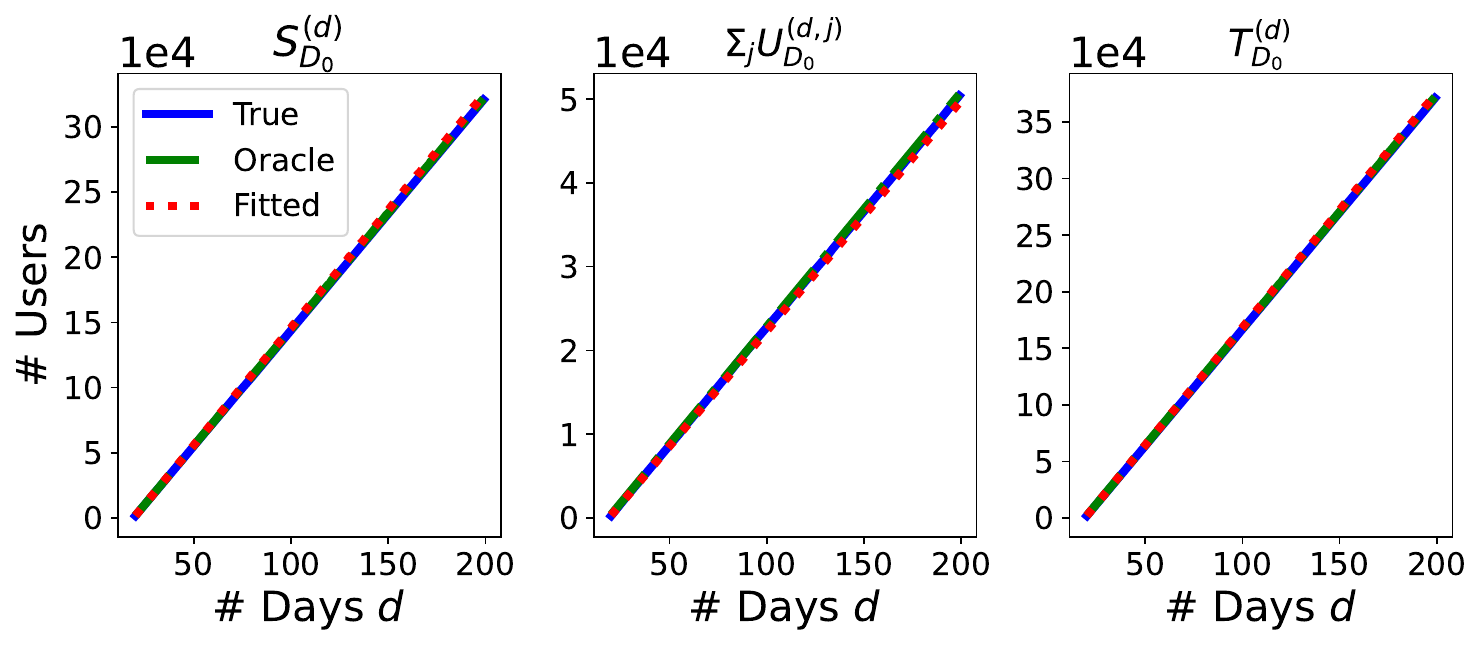}
    \caption{Prediction of future user activity (left, $S_{\pilotdays}^{(\dayidx)}$, center, $\sum_\freq \news{\pilotdays}{\dayidx, \freq}$, right $T_{\pilotdays}^{(\dayidx)}$). We plot the value of the statistic (vertical axis) as the number of days $\dayidx$ increases (horizontal axis). We compare the true value in future samples (solid blue line) to the predicted posterior mean using the fitted (red) and the true (green) hyperparameters.}
    \label{fig:synthetic_sums}
\end{figure}

We first draw synthetic data from the model \eqref{eq:model} using the sampling scheme of \Cref{sec:theory_sampling}. Over a large collection of hyperparameter values, and durations of the pilot and follow-up study $\pilotdays$ and $\followupdays$, we find that maximizing the marginal likelihood \eqref{eq:likelihood} slightly outperforms the regression approach outlined in \Cref{sec:inference}. Whenever possible, we therefore use this approach. When re-trigger information is not available (e.g., for the ASOS data of \citet{liu2021datasets},  \Cref{fig:accuracy_2}), we adopt the regression approach. We find that even for $\pilotdays\ll\followupdays$, we can form accurate predictions for $\news{\pilotdays}{\followupdays}$ and $\news{\pilotdays}{\followupdays, \freq}$ (\Cref{fig:synthetic_1,fig:synthetic_sums}). 

\begin{figure}[ht]
    \centering
    \includegraphics[width=\linewidth]{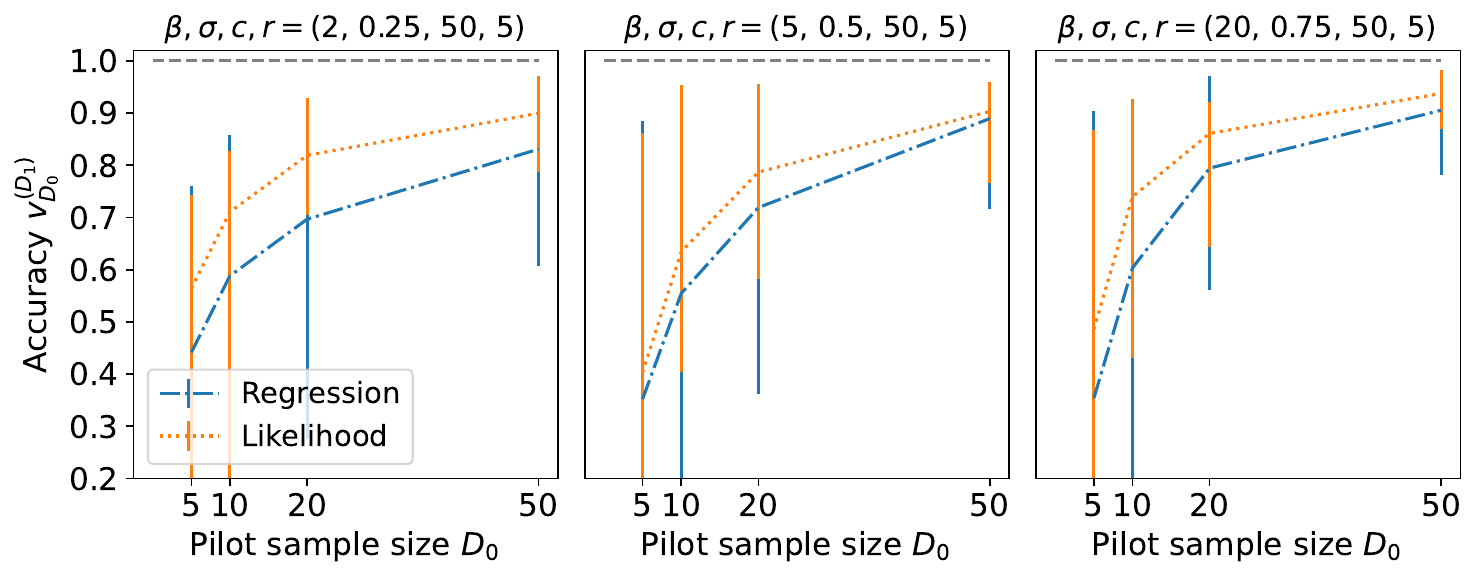}
    \caption{Prediction accuracy $v_{\pilotdays}^{(\followupdays)}$ of the Bayesian nonparametric predictor $\prednews{\pilotdays}{\followupdays}$ on data from the model.}
    \label{fig:synthetic_accuracy}
\end{figure}

We sample $D=200$ days of trigger data $\triggerdata{1:D}$ from \Cref{eq:model} with parameters $(\mass, \tail, \tilting, r) = (0.1, 0.5, 50, 5)$. We retain the first $\pilotdays=20$ days to fit the hyperparameters and form predictions for $\news{\pilotdays}{\dayidx}$ and $\news{\pilotdays}{\dayidx, \freq}$ by using the posterior mean induced by the fitted parameters, using the characterization provided in \Cref{sec:theory_prediction}. 
For all the statistics considered, we are able to form predictions to a high degree of accuracy. Further, we verify that when the data is drawn from the ``true'' model \eqref{eq:model}, even extremely small sample sizes allow us to form accurate predictions of future user activity. This is important, as experimenters typically want to form predictions in the early phases of the experiment. We show this in \Cref{fig:synthetic_accuracy}, where for a fixed total dimension of the sample $D = 500$ we draw $M=100$ samples from the model \eqref{eq:model}, and test both the log-likelihood and the regression approach to fit the hyperparameters. We report in \Cref{fig:synthetic_accuracy} the median accuracy $v_{\pilotdays}^{(\followupdays)}$ and a centred empirical 80\% interval as the size of the pilot $\pilotdays$ increases
for three different choices of the hyperparameters $\mass, \tail, \tilting, r$.

\paragraph{Synthetic data from Zipf distributions} 

\begin{figure}[ht]
    \centering
    \includegraphics[width=\linewidth]{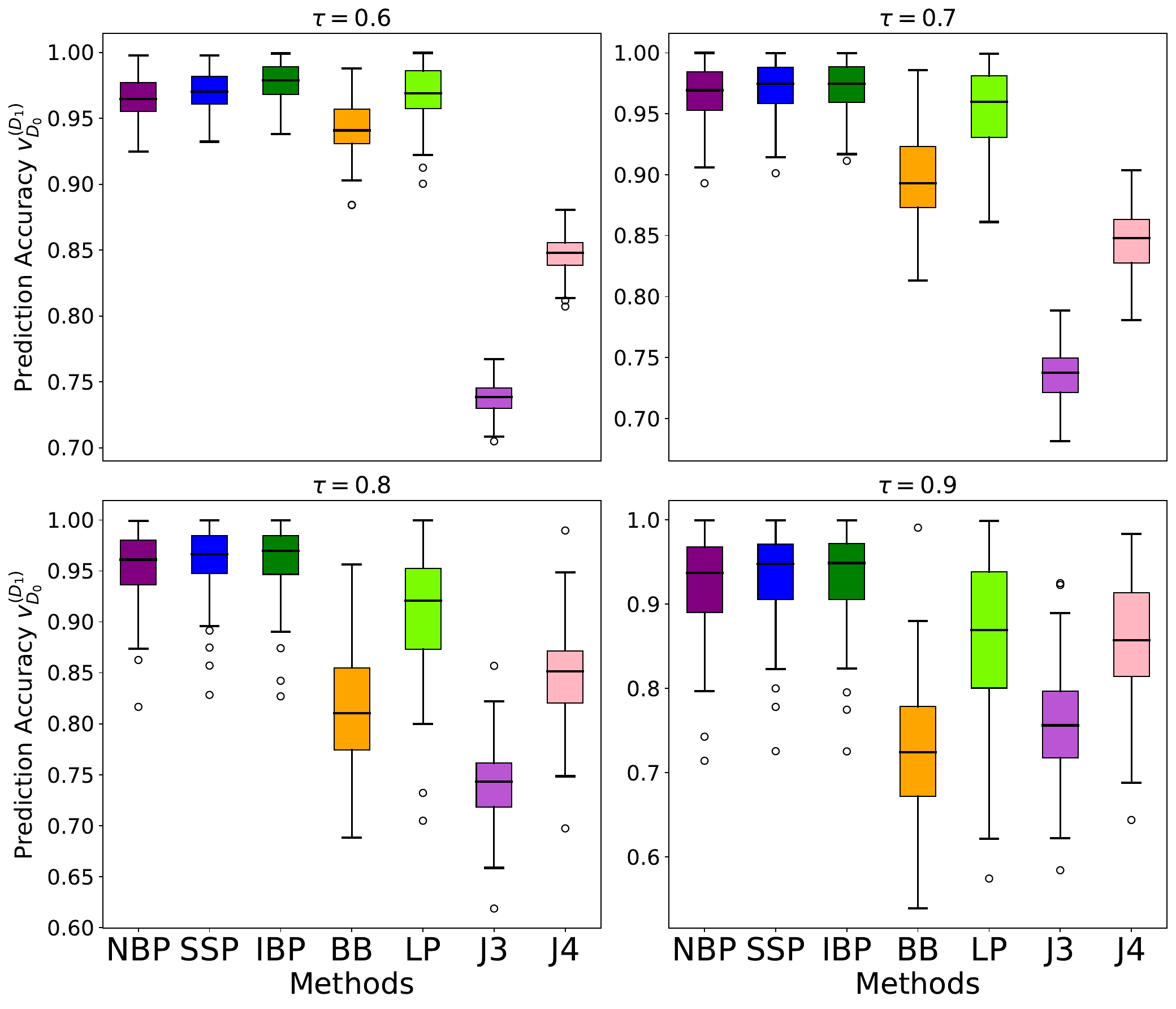}
    \caption{Prediction accuracy $v_{\pilotdays}^{(\followupdays)}$ of predictors $\prednews{\pilotdays}{\followupdays}$ on synthetic data from the Zipfian model for different choices of the parameter $\tau$. Here, $\pilotdays=10$ and $\followupdays=50$.}
    \label{fig:synthetic_accuracy_zipf}
\end{figure}

Next, we evaluate the performance of our predictors against that of competing methods on synthetic data from a Zipf-Poisson distribution. In particular, for a given cardinality $N_{\infty} = 1{,}000{,}000$ of possible users we draw trigger data as follows. We endow each user $\unitidx$ with a triggering rate $\theta_\unitidx = \unitidx^{-\tau}$, with $\tau = 0.6, 0.7, 0.8, 0.9$. Then, for every experimental day $\dayidx$, we determine whether each user $\unitidx$ re-triggers by flipping a Bernoulli coin, $X_{\dayidx, \unitidx} \sim \mathrm{Bernoulli}(\theta_\unitidx)$. Conditionally on $X_{\dayidx, \unitidx} = 1$, we sample $\triggerdata{\dayidx, \unitidx} \sim \mathrm{tPoisson}(1+m_{\dayidx-1, \unitidx}/\dayidx; 0)$, where $m_{\dayidx-1, \unitidx}$ is the total re-trigger rate of unit $\unitidx$ up to day $\dayidx-1$ and $\mathrm{tPoisson}(a;b)$ is the law of a Poisson random variable with parameter $a$ and supported on $\{b+1, b+2, \ldots \}$. For each value $\tau$, we generate $M=100$ datasets. We retain only $\pilotdays=5$ days for training, and compute the predictive accuracy $v_{\pilotdays}^{(\followupdays)}$ for $\followupdays = 50$ days ahead. Our results in \Cref{fig:synthetic_accuracy_zipf} show that across values of $\tau$ our method (NBP) is competitive with the best alternatives in the literature, achieving a predictive performance comparable to the scaled-stable-beta-Bernoulli process (SSP) and the Indian buffet process (IBP). We omit results for lower Jackknife estimators, Good-Toulmin, beta-geometric as they are significantly worse than other competitors. Each boxplot reports the median of the accuracy, together with a box of width given by the interquartile range.

\begin{figure}[ht]
    \centering
    \includegraphics[width=\linewidth]{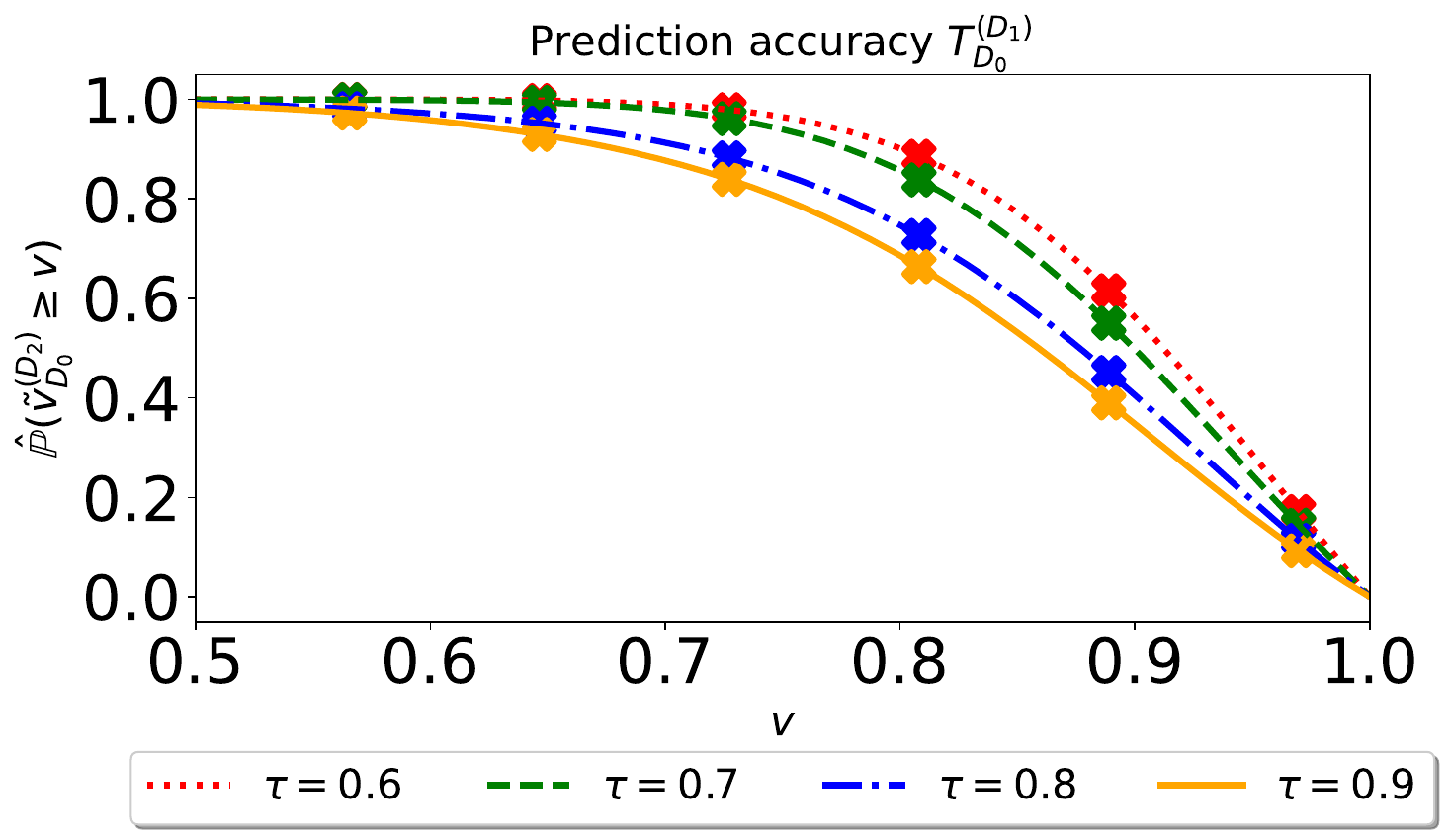}
    \caption{Prediction accuracy $\tilde{v}_{10}^{(50)}$ of the NBP predictor on Zipf data for different choices of the parameter $\tau$.
    }
    \label{fig:synthetic_accuracy_sums}
\end{figure}

Additionally, we show in \Cref{fig:synthetic_accuracy_sums} that the total sum predictor $\hat{T}_{\pilotdays}^{(\followupdays)}$ achieves great accuracy in predicting the total sum via a survival plot. While beyond the scope of the present paper, total re-trigger rates can be leveraged in forming better estimates of long-term treatment effects and to better predict the duration needed for an experiment to deliver significant results \citep{richardson2022bayesian,wan2023experimentation}.

\subsection{Experiments on real data}

\begin{figure}
    \centering
    \includegraphics[width=\linewidth]{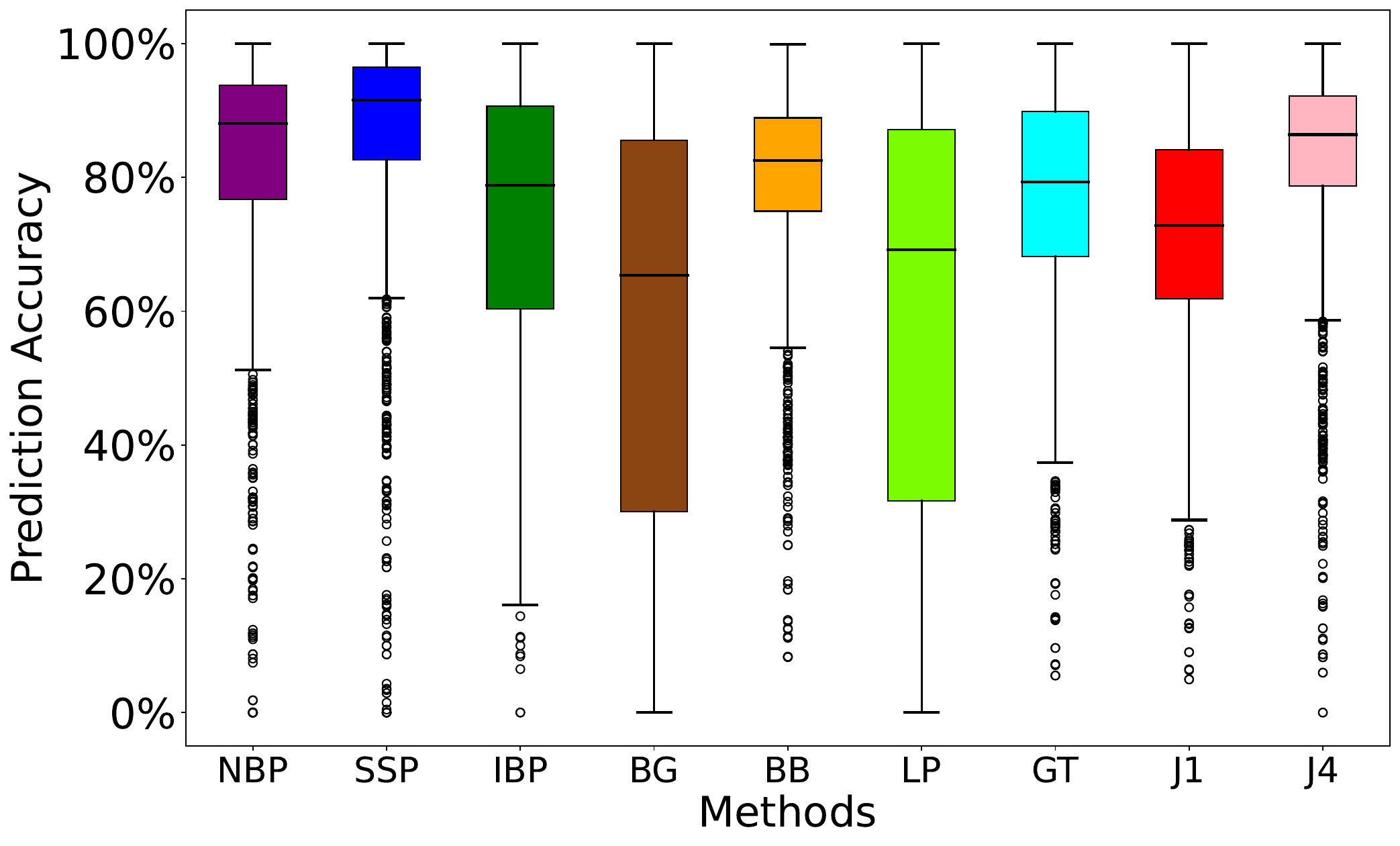}
    \caption{Prediction accuracy $v_{7}^{(21)}$ on proprietary data.}
    \label{fig:accuracy_1}
\end{figure}

\begin{figure}
    \centering
    \includegraphics[width=\linewidth]{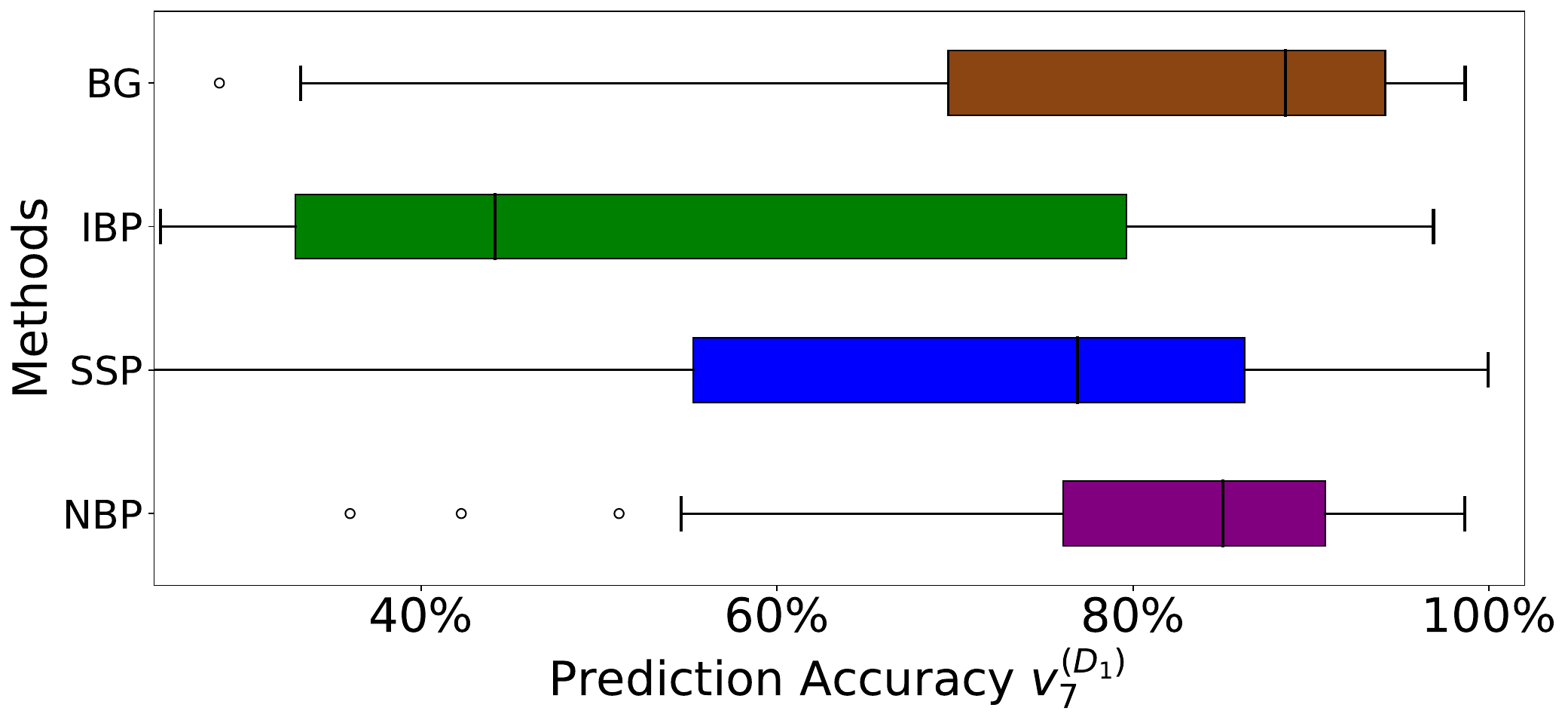}
    \caption{Prediction accuracy $v_{7}^{(\followupdays)}$ on ASOS data.}
    \label{fig:accuracy_2}
\end{figure}

Last, we assess the empirical performance of our and competing methods on real data. We use trigger data from a large meta analysis containing $1{,}774$ experiments run in production by a large technology company of varying size (from small experiments with a few thousand users, to large experiments with hundreds of million users per experiment). We also consider the ASOS data provided in \citet{liu2021datasets}, from which we retain a total of 76 treatment arms across experiments. For both meta analyses, we retain $\pilotdays=7$ days worth of pilot data, and predict for either $\pilotdays=21$ days ahead (proprietary data), or for the longest available duration (ASOS data) using our new method as well as competing methods. Results are provided in \Cref{fig:accuracy_1,fig:accuracy_2}. Our method is competitive with the best available alternatives in the literature. 
The ASOS data \citep{liu2021datasets} only provides for each experiment the number of distinct users first triggering on a given date, and not the number of users who retriggered a given number of times throughout the experiment,  information needed to fit other competing methods (see \Cref{sec:app_competing} for additional details). We therefore only consider the Bayesian models which can be trained with the available data. Again, the NBP is competitive with the best available alternative. 

\begin{figure}[ht!]
    \centering
    \includegraphics[width=\linewidth]{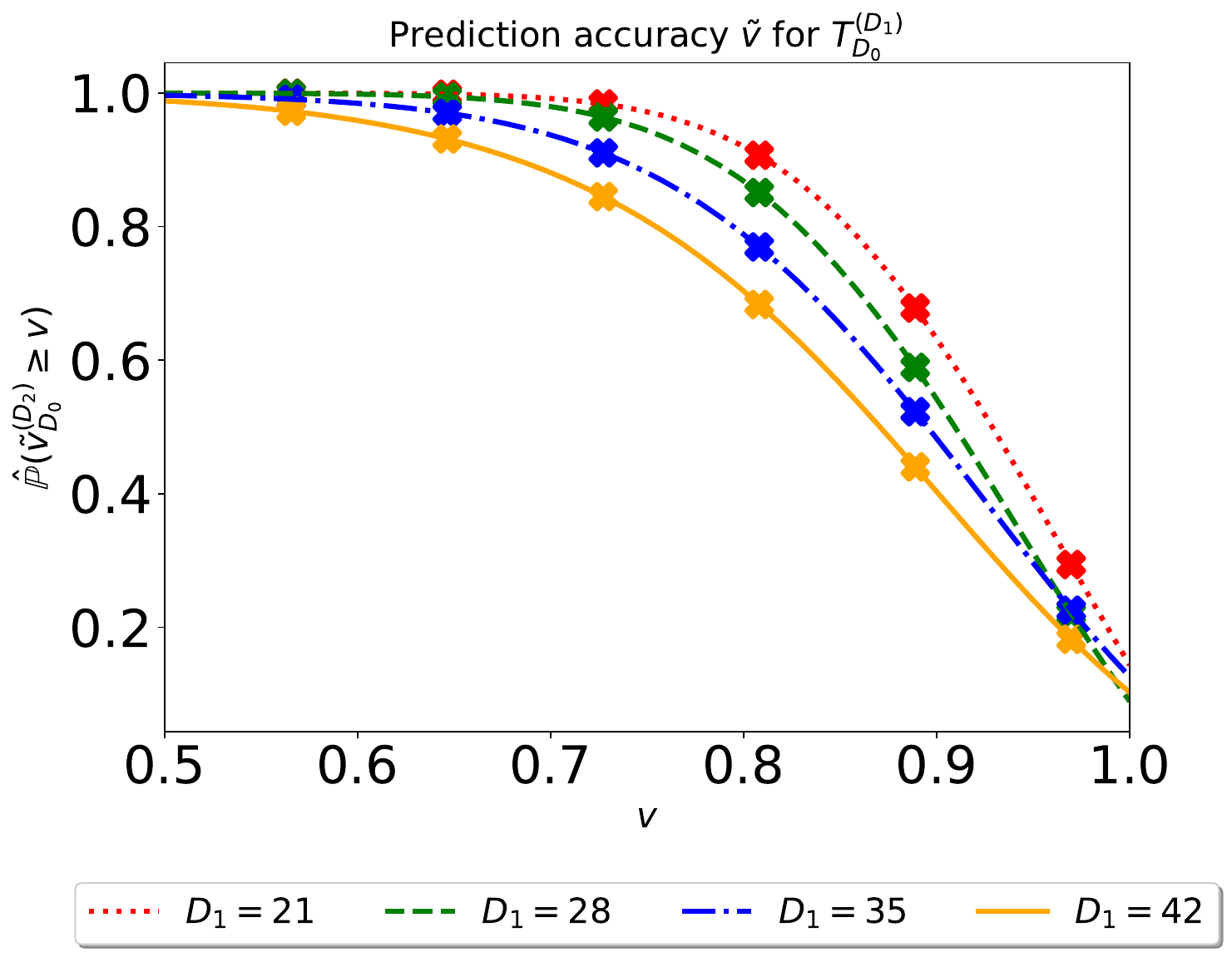}
    \caption{Survival plot for the prediction accuracy $\tilde{v}_{\pilotdays}^{(\followupdays)}$ of the NBP predictor for the total triggering activity on the proprietary data. For a given accuracy level (horizontal axis), we plot the fraction of experiments achieving at least that level of accuracy (vertical axis). Different lines refer to different extrapolation values $D_1$.
    }
    \label{fig:true_accuracy_sums}
\end{figure}

Last, we assess in \Cref{fig:true_accuracy_sums} the performance of our method on a smaller subset of 50 experiments at predicting the total future activity rate. We retain $\pilotdays=7$ days for training, and extrapolate at different horizons $\followupdays=14, 21, 28, 35$. For this task, no alternative method is available.
\if0\arxiv
\subsection{Reproducibility workflow}

We provide detailed and rigorous mathematical proofs of all the results presented in \Cref{sec:model} in \Cref{sec:proofs}. Additionally, as part of our supplementary material, we provide data and \texttt{python} code to reproduce all our analyses on simulated data, as well as the results for the publicly available ASOS data. Data to reproduce results on proprietary data is omitted. Details to generate results and reproduce plots in the main text and in the appendix are provided within the \texttt{README.md} file.
\fi
\section{Discussion}

In this paper, we have introduced a novel Bayesian nonparametric method which can be used to predict future user activity in online A/B tests, which complements relevant related research in the field \citep{richardson2022bayesian, wan2023experimentation}. 
To the best of our knowledge, our work is the first one linking Bayesian nonparametric models \citep{teh2009indian,masoero2022more} to the problem of sample size prediction in online A/B testing.
By making this connection, our work provides a set of theoretically well studied, simple and scalable algorithms. We show on extensive experiments on real and synthetic data that our newly proposed approach is competitive with the best existing alternatives when predicting the total number of future users that are going to be sampled in the experiment, and is easy to fit to the data. 
In contrast to existing methods in the literature, our method also allows prediction of the rate at which previously seen and  yet-to-be seen users are going to be observed in future experimental days.

\if0\arxiv
{
As part of our contribution, we provide extensively tested code that implements not only our methods, but also a suite of competing alternative probabilistic methods that can be used for the tasks at hand.
}
\fi

Our work can be seen as an extension of \citet{camerlenghi2022scaled}. However, it is crucial to note that it is not a mere extension: it introduces novel concepts and methodologies that significantly expand the existing research landscape, not only broadening the scope of the methodology proposed in \citet{camerlenghi2022scaled}, but also addressing practical challenges in online A/B testing that had not been previously explored. We believe that the application of this novel and relatively under-studied methodology to the  domain of online, large-scale causal inference opens avenues for future research in this space. 

We envision a number of exciting avenues for future work. In particular, while not investigated in the present work, we expect that predicting more granular information about future re-trigger rates can be used by experimenters to improve how they determine the optimal duration of an experiment \citep{wan2023experimentation}. Moreover, this information can be used to form better estimates of long-term-effects of treatments. This information can therefore substantially reduce experimentation cost, and improve decision making.
Additionally, it would be interesting to understand whether predictions of future trigger activity can be directly leveraged to increase the sensitivity of user engagement metrics in online A/B tests, an idea previously 
explored in \citet{drutsa2015future}.

\newpage
\newpage
\bibliographystyle{ACM-Reference-Format}
\bibliography{references}
\newpage
\appendix
\onecolumn
\section{Proofs} \label{sec:proofs}

We now present the proofs of the results presented in \Cref{sec:posterior}. We refer the interested reader to \citet{camerlenghi2022scaled} for a thorough discussion of scaled processes and their use as prior distributions in Bayesian nonparametric models.

Recall the definition of the L\'{e}vy measure $\nu$ given in \Cref{eq:levy}. Let $\lambda(s)$ be the corresponding L\'{e}vy density, i.e.
\[
    \lambda(s) = \tail s^{-1-\tail} \ind(s>0).
\]
Define the tilting function $h_{\mass, \tilting}$, for $\mass>0$ and $\tilting>0$
\[
    h_{\mass, \tilting}(s) = \frac{\mass^{\tilting+1}}{\Gamma(\tilting+1)} s^{-\tilting \tail} \exp\left\{ - (\mass-1) s^{-\tail} \right\}.
\]
Additionally, let
\begin{align}
    \phi_d(\eta) &= \int_0^1 \left[1-(1-s)^r\right](1-s)^{r(d-1)} \eta \lambda(\eta s) \de s \nonumber \\
    &= \tail \eta^{-\tail} \int_0^1 \left[1-(1-s)^r\right](1-s)^{r(d-1)}  s^{-1-\tail} \de s \nonumber \\
    &= \tail \eta^{-\tail} \left\{ \int_0^1 (1-s)^{r(d-1)}  s^{-1-\tail} \de s  -  \int_0^1 (1-s)^{rd}  s^{-1-\tail} \de s \right\} \nonumber \\
    &= \tail \eta^{-\tail} \left[B(r(d-1)+1,-\tail) - B(rd+1,-\tail)\right]. \nonumber
\end{align}
\subsection{Proof of Proposition \ref{prop:jump}}

\begin{proof}
    The (prior) distribution of the tilted largest jump $\Delta_{1,h_{\mass,\tilting}}$ of the $\SSP(\nu, h_{\mass,\tilting})$ process is given by 
\begin{equation}
    f_{\Delta_{1, h_{\mass,\tilting}}}(\eta) = \frac{\tail \mass^{\tilting+1}}{\Gamma(\tilting+1)} \eta^{-\tail(\tilting+1)-1}\exp\left\{ - \mass \eta^{-\tail} \right\} \ind(\eta>0). \nonumber
\end{equation}
See \citet[C.2]{camerlenghi2022scaled}.

Following the proof of \citet[Proposition 7]{camerlenghi2022scaled}, denote by $\Delta_{1,h_{\mass,\tilting}}\mid \triggerdata{1:\pilotdays}$ the largest jump of the SSP prior conditionally given $\pilotdays$ draws $\triggerdata{1:\pilotdays}$. Its distribution is proportional to:
\begin{align}
    f_{\Delta_{1,h_{\mass,\tilting}}\mid \triggerdata{1:\pilotdays}}(\eta) &\propto f_{\Delta_{1,h_{\mass,\tilting}}}(\eta) \pr(Z_{1:\pilotdays} \mid \Delta_{1,h_{\mass,\tilting}} = \eta) \nonumber\\
    & \propto \exp\left\{ - \sum_{\dayidx=1}^{\pilotdays} \phi_d(\eta) - \mass \eta^{-\tail} \right\} \eta^{-\tail(\unittotal+\tilting+1)-1} \ind(\eta\in\RE_+)\nonumber\\
    &= \exp\left\{ - \eta^{-\tail} \left[\mass + \psi_0^{({\pilotdays})}\right] \right\} \eta^{-\tail(\unittotal+\tilting+1)-1}\ind(\eta\in\RE_+).\nonumber
\end{align}
That is, $\Delta_{1,h_{\mass,\tilting}}^{-\tail}\mid \triggerdata{1:\pilotdays} \sim \mathrm{Gamma}(\unittotal+\tilting+1, \mass + \psi_0^{(\pilotdays)})$. That is to say, the posterior density of $\Delta_{1,h_{\mass,\tilting}} \mid \triggerdata{1:\pilotdays}$ is given by:
\[
    f_{\Delta_{1,h_{\mass,\tilting}}\mid \triggerdata{1:\pilotdays}}(\eta) = \frac{\tail (\mass + \psi_0^{({\pilotdays})})^{\unittotal+\tilting+1}}{\Gamma(\unittotal+\tilting+1)} \eta^{-\tail(\unittotal+\tilting+1)-1}
    \exp\left\{ - \eta^{-\tail}(\beta+\psi_0^{(\pilotdays)}) \right\} \ind(\eta \in \RE_+).
\]
\end{proof}

\subsection{Proof of Proposition \ref{prop:posterior}}

\begin{proof}
Following \citet[Proposition 6]{camerlenghi2022scaled}, conditionally on (i) the value of the tilted largest jump $\Delta_{1,h_{\mass,\tilting}}$ and (ii) the sample $\triggerdata{1:\pilotdays}$, the measure $\randommeasure$ is a completely random measure. Its posterior representation consists of the sum of an ``ordinary'' component $\tilde{\Theta}$, and a ``fixed-location'' component \citep{broderick2018posterior}, and is given by
\[
    \randommeasure \mid \Delta_{1,h_{\mass,\tilting}}, \triggerdata{1:\pilotdays} \overset{d}{=} \tilde{\randommeasure} + \sum_{n=1}^{\unittotal_{\pilotdays}} J_\unitidx \delta_{\genericunit}(\cdot),
\]
where
\begin{itemize}
    \item The ordinary component is 
    \[
        \tilde{\randommeasure} \mid \Delta_{1,h_{\mass,\tilting}} \sim \CRM(\tilde{\nu}(\de s) \times P_0(\de \omega))
    \]
    with
    \[
        \tilde{\nu}(\de s) = (1-s)^{r\pilotdays}\Delta_{1,h_{\mass,\tilting}}\lambda(\Delta_{1,h_{\mass,\tilting}}s)  \ind(s\in(0,1))\de s.
    \]
    \item Recall the definition of $m_{{\pilotdays},\unitidx} := \sum_{\dayidx=1}^D \triggerdata{\dayidx, \unitidx}$. The jumps of the fixed-location component $J_{1:\unittotal_{\pilotdays}}$ are independent and have distribution proportional to
    \begin{align}
        f_{J_\unitidx \mid \Delta_{1,h_{\mass,\tilting}}}(s) & \propto (1-s)^{r(D-m_{D,\unitidx})}(1-(1-s)^r)^{m_{{\pilotdays},\unitidx}} \prod_{\dayidx:a_{\dayidx, \unitidx}>0}^{\pilotdays} \left[ \frac{(1-s)^r s^{a_{\dayidx, \unitidx}}}{1-(1-s)^r}  \right] s^{-1-\tail} \nonumber \\
        & \propto (1-s)^{r\pilotdays+1-1}s^{m_{{\pilotdays},\unitidx}-\tail-1} \\
        &\overset{d}{=} \betad(m_{{\pilotdays},\unitidx}-\tail, r\pilotdays+1). \label{eq:proof_posterior_jumps}
    \end{align}
\end{itemize}
\end{proof}

\subsection{Proof of Proposition \ref{prop:predictive}}

\begin{proof}
    Following \citet[Propostion 3]{camerlenghi2022scaled}, the conditional (predictive) distribution of $\triggerdata{\pilotdays+1}$ given $\triggerdata{1:\pilotdays}$ and $\Delta_{1, h_{\mass, \tilting}}$ coincides with the distribution of 
    \begin{equation}
       \triggerdata{\pilotdays+1} \mid \triggerdata{1:\pilotdays}, \Delta_{1, h_{\mass, \tilting}} \overset{d}{=} \triggerdata{\pilotdays+1}' + \sum_{n=1}^{\unittotal_{\pilotdays}} A_{\pilotdays+1, \unitidx} \delta_{W_\unitidx^\star},  \nonumber
    \end{equation}
    
    where
    \begin{enumerate}
        \item $\triggerdata{\pilotdays+1}' \mid \tilde{\randommeasure}, \Delta_{1, h_{\mass, \tilting}}  = \sum_{\unitidx\ge 1} A'_{\pilotdays+1,\unitidx}\delta_{W'_\unitidx} \sim \NBP(\tilde{\randommeasure})$.
        \item The random variables $A_{D+1,\unitidx} \overset{\text{ind}}{\sim} \NegBin{r}{J_\unitidx}$, with $J_\unitidx$ distributed as in \Cref{eq:proof_posterior_jumps}.
    \end{enumerate}
\end{proof}

\subsection{Proof of Proposition \ref{prop:marg}}

\begin{proof}
For the marginal distribution, recall that under the negative binomial model of \Cref{eq:model}, conditionally on $\randommeasure$, the probability of customer $\unitidx$ not triggering on any given day $\dayidx$ is given by $(1-\theta_\unitidx)^r$, where $\theta_\unitidx$ is the $\unitidx$-th scaled jump of the prior.
Then following \citet[Proposition B2.5]{camerlenghi2022scaled}, the marginal distribution of $\triggerdata{1:\pilotdays}$ is given by
\begin{align}
    \pr(\triggerdata{1:\pilotdays}) 
    &=\int_0^{\infty}  e^{-\sum_{\dayidx=1}^{\pilotdays} \phi_\dayidx(\eta)} \prod_{\unitidx=1}^\unittotal \left\{ \tail \eta^{-\tail} \int_0^1 \left[1-(1-s)^r\right]^{m_{\pilotdays,\unitidx}} (1-s)^{(\pilotdays-m_{{\pilotdays},\unitidx})r} s^{-1-\tail}  \right. \nonumber \\
    &\quad \left. \prod_{\dayidx:a_{\dayidx, \unitidx}>0}^{\pilotdays} \left[ \frac{\binom{a_{\dayidx, \unitidx} + r - 1}{a_{\dayidx, \unitidx}}}{1-(1-s)^r} (1-s)^r s^{a_{\dayidx, \unitidx}} \right] \de s \right\} f_{\Delta_{1,f_{\tilting, \mass}}}(\eta) \de \eta. \nonumber
\end{align}
Notice
\[
    \exp\left\{-\sum_{\dayidx=1}^{\pilotdays} \phi_\dayidx(\eta)\right\} = \exp\left\{-\tail \eta^{-\tail} \left[B(1,-\tail) - B(r\pilotdays+1,-\tail)\right]\right\}.
\]
Moreover,
\begin{align}
    \int_0^1& \left[1-(1-s)^r\right]^{m_\unitidx} (1-s)^{({\pilotdays}-m_\unitidx)r} s^{-1-\tail} \prod_{\dayidx:a_{\dayidx, \unitidx}>0}^D \left[ \frac{\binom{a_{\dayidx, \unitidx} + r - 1}{a_{\dayidx, \unitidx}}}{1-(1-s)^r} (1-s)^r s^{a_{\dayidx, \unitidx}} \right] \de s  \nonumber \\
    = &\prod_{\dayidx:a_{\dayidx, \unitidx}>0}^{\pilotdays} \left[ \binom{a_{\dayidx, \unitidx} + r - 1}{a_{\dayidx, \unitidx}} \right]  \int_0^1 (1-s)^{r{\pilotdays}} s^{m_{\unitidx}-\tail-1} \de s \nonumber \\
    &= \prod_{\dayidx:a_{\dayidx, \unitidx}>0}^{\pilotdays} \left[ \binom{a_{\dayidx, \unitidx} + r - 1}{a_{\dayidx, \unitidx}}\right] B(r{\pilotdays}+1, m_\unitidx-\tail+1). \nonumber
\end{align}
Hence,
\begin{align}
    \pr(\triggerdata{1:\pilotdays}) 
    &= \frac{\tail^{\unittotal+1}\mass^{\tilting+1}}{\Gamma(\tilting+1)}\prod_{\unitidx=1}^{\unittotal}\left\{\prod_{\dayidx:a_{\dayidx, \unitidx}>0}^{\pilotdays} \left[ \binom{a_{\dayidx, \unitidx} + r - 1}{a_{\dayidx, \unitidx}}\right] B(r\pilotdays+1, m_\unitidx-\tail+1)\right\} \nonumber \\
    &\int_0^\infty \exp\left\{-\eta^{-\tail} \left(\tail\left[B(1,-\tail) - B(r\pilotdays+1,-\tail)\right] + \mass\right) \right\} \eta^{- \tail (\unittotal + \tilting + 1) - 1 } \de \eta \nonumber \\
    &= \frac{\tail^{\unittotal+1}\mass^{\tilting+1}}{\Gamma(\tilting+1)}\prod_{\unitidx=1}^{\unittotal}\left\{\prod_{\dayidx:a_{\dayidx, \unitidx}>0}^{\pilotdays} \left[ \binom{a_{\dayidx, \unitidx} + r - 1}{a_{\dayidx, \unitidx}}\right] B(r\pilotdays+1, m_\unitidx-\tail+1) \right\} \nonumber \\
    &\quad \times\; \frac{\Gamma(\unittotal+\tilting+1)}{\tail \left(\mass + \tail\left[B(1,-\tail) - B(r\pilotdays+1,-\tail)\right]\right)^{\unittotal+\tilting+1}} \nonumber \\
    &= \frac{\tail^{\unittotal}\mass^{\tilting+1}}{\left(\mass + \psi_0^{({\pilotdays})} \right)^{\unittotal+\tilting+1}} \frac{\Gamma(\unittotal+\tilting+1)}{\Gamma(c)}
    \prod_{\unitidx=1}^{\unittotal} \xi(a_{1:{\pilotdays}, \unitidx}, r, \tail), \nonumber
\end{align}
with 
\[
    \psi_{x}^{(y)}= \tail\left[B(rx+1,-\tail) - B(r(x+y)+1,-\tail)\right]
\]
so that
\[
    \psi_0^{(\pilotdays)}= \tail\left[B(1,-\tail) - B(r\pilotdays+1,-\tail)\right],
\]
and 
\[
    \xi(a_{1:\pilotdays, \unitidx}, r, \tail):=\prod_{\dayidx:a_{\dayidx, \unitidx}>0}^{\pilotdays} \left[ \binom{a_{\dayidx, \unitidx} + r - 1}{a_{\dayidx, \unitidx}}\right] B(r\pilotdays+1, m_\unitidx-\tail+1).
\]
\end{proof}

\subsection{Proof of Proposition \ref{prop:new_users}}
\begin{proof}
    The distribution of the number of new customers that are going to be observed for the first time between day $\pilotdays$ and day $\followupdays$, $\news{\pilotdays}{\followupdays}$ can be derived following the argument in \citet[Theorem 2]{camerlenghi2022scaled}. Notice that the distribution of the new customers conditionally on the largest jump must be a Poisson distribution from the properties of CRMs. We can write from the predictive representation in \Cref{{eq:predictive_characterization}}
\[
    \news{\pilotdays}{\followupdays} \mid \triggerdata{1:\pilotdays}, \Delta_{1, h_{\mass, \tilting}} \overset{d}{=} \sum_{\unitidx\ge 1} \ind\left( \sum_{\dayidx=1}^{\followupdays} \ind\left(A'_{\pilotdays+\dayidx, \unitidx} > 0\right)>0 \right).
\]
Here $\ind(A'_{\pilotdays+\dayidx, \unitidx}>0)$ is --- conditionally on $\tilde{\randommeasure}$ --- a Bernoulli random variable with parameter $1-(1-\tilde{\theta}_\unitidx)^r$, where $\tilde{\theta}_\unitidx$ are the jumps of a Poisson point process with L{\'e}vy intensity
\[
    \tilde{\lambda}(s)\de s = \tail \Delta_{1, h_{\mass, \tilting}}^{-\tail} (1-s)^{r\pilotdays} s^{-1-\tail}  \ind(s \in (0,1)) \de s.
\]
It follows that 
\begin{align}
    \EE\left[ t^{\news{\pilotdays}{\followupdays}} \mid \triggerdata{1:\pilotdays}, \tilde{\randommeasure} \right] & = \EE\left[ \prod_{\unitidx\ge 1} \left\{ (t-1) \left( 1 - \prod_{\dayidx=1}^{\followupdays} \pr(A'_{\pilotdays+\dayidx, \unitidx} = 0 \mid \tilde{\randommeasure} \right) + 1 \right\} \right] \nonumber \\
    &= \EE\left[ \prod_{\unitidx\ge 1} \left\{ (t-1)\left[ 1-(1-\tilde{\theta}_\unitidx)^{r\followupdays} + 1 \right] \right\} \right] \nonumber \\
    &= \EE\left[ \exp\left\{ \sum_{\unitidx\ge 1} \log\left( (t-1)[1-(1-\tilde{\theta}_\unitidx]^{r\followupdays} \right) + 1 \right\} \right] \nonumber \\
    &= \exp\left\{ - (1-t) \int_0^1 (1-(1-s)^{r\followupdays})(1-s)^{r\pilotdays}  \tail \Delta_{1, h_{\mass, \tilting}}^{-\tail}  s^{-1-\tail} \de s\right\} \nonumber \\
    &= \exp\left\{ - (1-t) \Delta_{1, h_{\mass, \tilting}}^{-\tail} \tail [B(r\pilotdays+1, -\tail) - B(r(\pilotdays+\followupdays) +1, -\tail)\right\} \nonumber \\
    &= \exp\left\{ - (1-t) \Delta_{1, h_{\mass, \tilting}}^{-\tail} \psi_{\pilotdays}^{(\followupdays)} \right\}. \nonumber
\end{align}
One can then integrate with respect to the posterior distribution of $\Delta_{1, h_{\mass, \tilting}}^{-\tail}$ to obtain the final result:
\begin{align}
    \EE\left[ t^{\news{\pilotdays}{\followupdays}} \mid \triggerdata{1:\pilotdays}\right] &= 
    \frac{ (\mass + \psi_{\pilotdays}^{(\followupdays)} )^{\unittotal+\tilting+1}}{\Gamma(\unittotal+\tilting+1)} 
    \int_0^{\infty} \exp\left\{ - x[(1-t)\psi_{\pilotdays}^{(\followupdays)} + \mass + \psi_{\pilotdays}^{(\followupdays)}] \right\} x ^{\unittotal+\tilting}\de x \nonumber \\
    &= \left( \frac{\mass + \psi_{\pilotdays}^{(\followupdays)}}{\mass + \psi_{\pilotdays}^{(\pilotdays+\followupdays)} - t \psi_{\pilotdays}^{(\followupdays)}} \right)^{\unittotal+\tilting+1} \nonumber \\ 
    &= \left( \frac{1-p_{\pilotdays}^{(\followupdays)}}{1-t p_{\pilotdays}^{(\followupdays)}} \right)^{\unittotal+\tilting+1}, \nonumber
\end{align}
for any $t<1/|p_{\pilotdays}^{(\followupdays)}|$, with 
\[
    p_{\pilotdays}^{(\followupdays)}:=\frac{\psi_{\pilotdays}^{(\followupdays)}}{\mass + \psi_{0}^{(\pilotdays+\followupdays)}} \le 1.
\]
This is the probability generating function of a negative binomial distribution with success rate $p_{\pilotdays}^{(\followupdays)}$ and $\unittotal+\tilting+1$ failures:
\[
    \pr(\news{\pilotdays}{\followupdays} = \ell \mid \triggerdata{1:\pilotdays}) = \binom{\ell + \unittotal+\tilting}{\ell} (p_{\pilotdays}^{(\followupdays)})^{\ell} (1-p_{\pilotdays}^{(\followupdays)})^r \ind\{\ell \in \NN\}.
\]

\end{proof}

\subsection{Proof of Proposition \ref{prop:new_retrigger}}

\begin{proof}

The distribution of the number of new customers that are going to be observed for the first time between day $\pilotdays$ and day $\followupdays$ with a given total frequency $\freq$, $\news{\pilotdays}{\followupdays, \freq}$ can be derived following the argument in \citet[Theorem 2]{camerlenghi2022scaled}. Notice that the distribution of the new customers conditionally on the largest jump must be a Poisson distribution from the properties of CRMs. We can write from the predictive representation in \Cref{{eq:predictive_characterization}}
\[
    \news{\pilotdays}{\followupdays, \freq} \mid \triggerdata{1:\pilotdays}, \Delta_{1, h_{\mass, \tilting}} \overset{d}{=} \sum_{\unitidx\ge 1} \ind\left( \sum_{\dayidx=1}^{\followupdays} A'_{\pilotdays+\dayidx, \unitidx} = \freq \right).
\]
Here $A'_{\pilotdays+\dayidx, \unitidx}$ is --- conditionally on $\tilde{\randommeasure}$ --- a negative binomial random variable with parameters $r$, $\tilde{\theta}_\unitidx$, where $\tilde{\theta}_\unitidx$ are the jumps of a Poisson point process with L{\'e}vy intensity
\[
    \tilde{\lambda}(s)\de s = \tail \Delta_{1, h_{\mass, \tilting}}^{-\tail} (1-s)^{r\pilotdays} s^{-1-\tail}  \ind(s \in (0,1)) \de s.
\]
Now observing that the random variable
\[
    S_{\pilotdays, \followupdays, \unitidx}:= \sum_{\dayidx=1}^{\followupdays} A'_{\pilotdays+d, \unitidx}
\]
is a sum of i.i.d.\ negative binomial random variables with parameters $r, \tilde{\theta}_\unitidx$ conditionally on $\tilde{\randommeasure}$, it holds that $S_{\pilotdays, \followupdays, \unitidx}\mid \tilde{\randommeasure} \sim \NegBin{\followupdays r}{\tilde{\theta}_\unitidx}$.
It follows that 
\begin{align*}
    \EE&\left[ t^{\news{\pilotdays}{\followupdays, \freq}}  \mid \triggerdata{1:\pilotdays}, \tilde{\randommeasure} \right]  = \EE\left[ \EE\left[\prod_{\unitidx\ge 1} \left\{ (t-1) \ind \left( \sum_{\dayidx=1}^{\followupdays} A'_{\pilotdays+d, \unitidx} = \freq \right) + 1 \right\} \mid \tilde{\randommeasure} \right]\right] \\
    &= \EE\left[ \prod_{\unitidx\ge 1} \left\{ (t-1) \pr \left( S_{\pilotdays, \followupdays, \unitidx} = \freq \right) + 1 \right\}  \right] \\
    &= \EE\left[ \prod_{\unitidx\ge 1} \left\{ (t-1) \binom{\freq+r\followupdays +1}{\freq} (\tilde{\theta}_\unitidx)^\freq (1-\tilde{\theta}_\unitidx)^{r\followupdays} + 1 \right\}  \right] \\
    &= \EE\left[\exp\left\{ \sum_{\unitidx\ge 1} \log\left\{ (t-1) \binom{\freq+r\followupdays +1}{\freq} (\tilde{\theta}_\unitidx)^\freq (1-\tilde{\theta}_\unitidx)^{r\followupdays} + 1 \right\} \right\}  \right] \\
    &= \exp\left\{ - (1-t) \Delta_{1, h_{\mass, \tilting}}^{-\tail}  \binom{\freq+r\followupdays +1}{\freq} \tail \int  (1-s)^{r(\pilotdays+\followupdays)} s^{\freq-\tail-1}  \de s \right\}   \\
    &= \exp\left\{ - (1-t) \Delta_{1, h_{\mass, \tilting}}^{-\tail}  \binom{\freq+r\followupdays +1}{\freq}  \tail B\left[r(\pilotdays+\followupdays) +1, \freq - \tail \right] \right\}   \\
    &= \exp\left\{ - (1-t) \Delta_{1, h_{\mass, \tilting}}^{-\tail}  \rho_{\pilotdays}^{(\followupdays, \freq)} \right\},  
\end{align*}
with $\rho_{\pilotdays}^{(\followupdays, \freq)}:=\binom{\freq+r\followupdays +1}{\freq}  \tail B\left[r(\pilotdays+\followupdays) +1, \freq - \tail \right] $.
One can then integrate with respect to the posterior distribution of $\Delta_{1, h_{\mass, \tilting}}^{-\tail}$ to obtain the final result: 
\begin{align*}
    \EE\left[ t^{\news{\pilotdays}{\followupdays, \freq}} \mid \triggerdata{1:\pilotdays}\right] &= 
    \frac{ (\mass + \psi_{0}^{(\pilotdays)} )^{\unittotal+\tilting+1}}{\Gamma(\unittotal+\tilting+1)} 
    \int_0^{\infty} \exp\left\{ - x(1-t)\rho_{\pilotdays}^{(\followupdays, \freq)} \right\} x^{\unittotal+\tilting}\de x \\
    &= \left( \frac{\mass + \psi_{0}^{(\pilotdays)}}{\mass + \psi_{0}^{(\pilotdays)} + \rho_{\pilotdays}^{(\followupdays, \freq)} -t\rho_{\pilotdays}^{(\followupdays, \freq)}} \right)^{\unittotal+\tilting+1}  \\
    &= \left( \frac{1-p_{\pilotdays}^{(\followupdays, \freq)}}{1-tp_{\pilotdays}^{(\followupdays, \freq)}} \right)^{\unittotal+\tilting+1}
\end{align*}
for any $t<1/|p_{\pilotdays}^{(\followupdays, \freq)}|$, with 
\[
    p_{\pilotdays}^{(\followupdays, \freq)}:=\frac{\rho_{\pilotdays}^{(\followupdays,\freq)}}{\mass + \psi_{0}^{(\pilotdays)}+\rho_{\pilotdays}^{(\followupdays, \freq)}} \le 1.
\]
This is the probability generating function of a negative binomial distribution with success rate $p_{\pilotdays}^{(\followupdays, \freq)}$ and $\unittotal+\tilting+1$ failures:
\[
    \pr(\news{\pilotdays}{\followupdays} = \ell \mid \triggerdata{1:\pilotdays}) = \binom{\ell + \unittotal+\tilting}{\ell} (p_{\pilotdays}^{(\followupdays, \freq)})^{\ell} (1-p_{\pilotdays}^{(\followupdays, \freq)})^r \ind\{\ell \in \NN\}.
\]
\end{proof}

\subsection{Proof of Corollary \ref{cor:old_retrigger}}

The proof follows directly from the posterior representation provided in \Cref{prop:posterior}.

\subsection{Proof of Corollary \ref{cor:total_retrigger}}

The proof follows directly from combining \Cref{prop:new_retrigger,cor:old_retrigger}.

\subsection{Urn scheme representation for the data}

We here extend the urn scheme already discussed in \Cref{sec:theory_sampling} by providing ``conditional'' formulae for trigger counts of new and old users. These conditional formulae can be simpler to work with from a numerical standpoint.

\begin{enumerate}
    \item Sample $\news{\dayidx}{1} \mid Z_{1:\dayidx} \sim \NegBin{\unittotal_{\dayidx}+\tilting+1}{p_{\dayidx}^{(1)}}$ new users, as per \Cref{prop:new_users}.
    \item For each new user $\unitidx = \unittotal_{\dayidx} + 1,\ldots, \unittotal_{\dayidx} + \news{\dayidx}{1}$, sample the corresponding re-trigger count $A_{\dayidx+1, \unittotal_{\dayidx}+\unitidx}$ i.i.d.\ from the probability mass functions
    \[
        \pr(A = \ell) \propto\frac{\Gamma(\ell+r-1)}{\Gamma(\ell)} B(\ell-\tail, r\dayidx+1) \ind_{\{1,2,\ldots\}}(\ell).
    \]
    Alternatively, first sample $\news{\dayidx}{1}$ jumps $\tau_{\unittotal_{\dayidx}+\unitidx}$ for $\unitidx = 1, \ldots, \news{\dayidx}{1}$ i.i.d.\ from $\tau_{\unittotal_{\dayidx}+\unitidx} \iid f_{\dayidx}(s)$ with
    \[
        f_{\dayidx}(s)\propto (1-s)^{r\dayidx}(1-(1-s)^r) s^{-1-\tail} \ind (s\in[0,1])
    \]
    and then sample the re-trigger counts from a negative binomial distribution left truncated at $0$:
    \[
        A_{\dayidx+1, \unittotal_{\dayidx}+\unitidx} \mid \tau_{\unittotal_{\dayidx}+\unitidx} \overset{\mathrm{ind}}{\sim} \tNegBin{r}{\tau_{\unittotal_{\dayidx}+\unitidx}}{0}.
    \]
    \item For users $\unitidx=1,\ldots,\unittotal_{\dayidx}$ who already triggered in earlier days, sample counts from the marginal distribution
    \begin{equation*}
        \pr(A_{\dayidx+1, \unitidx}= \ell)\propto \binom{\ell + r - 1}{\ell} B(\ell+m_{\dayidx,\unitidx}-\tail, r\dayidx+1).
    \end{equation*}
    Equivalently from the conditional representation given by the hierarchical model
    \begin{align*}
        J_\unitidx  &\sim \betad(m_\unitidx - \tail, r\dayidx+1) \\
        A_{\dayidx+1, \unitidx} \mid J_\unitidx &\sim \NegBin{r}{J_\unitidx}.
    \end{align*}
\end{enumerate}
\section{Details on competing methods} \label{sec:app_competing}

To benchmark the performance of our newly proposed method, we consider a number of alternatives which have previously been proposed in the literature. We here provide additional details on these methods.

\subsection{Scaled-stable-beta-Bernoulli process (SSP) and Indian buffet process (IBP) predictors}

 The SSP and IBP are nonparametric Bayesian models similar to the NBP. The main difference, is that they are built for binary-valued observations. In a nutshell, for both models the construction is similar to the one we presented in \Cref{sec:model}, but assumes that we only measure whether a unit $\unitidx$ triggers on a given day $\dayidx$ or not, and not the number of times a unit re-trigger in the experiment of a given day. I.e.\ $\triggerdata{1:D} \in \{0,1\}^{D \times \unittotal_D}$. Practically this is achieved by letting the likelihood process in the construction $\LP(\ell, \randommeasure)$ be a Bernoulli process $\mathrm{BeP}(\randommeasure)$, i.e.
\[
    \triggerdata{\dayidx, \unitidx} \mid \randommeasure \sim \mathrm{Bernoulli} (\theta_{\unitidx}),
\]
i.i.d.\ across days $\dayidx$ for the same unit $\unitidx$, and independently across different units. 

The scaled-stable-beta-Bernoulli process (SSP) \citep{camerlenghi2022scaled} uses the same prior $\randommeasure$ that we have adopted here. The Indian buffet process uses a three-parameter beta process \citep{teh2009indian} as the prior for the frequencies $\randommeasure$. 

For both methods we use the induced predictive structure to obtain estimators of the quantities of interest, mirroring our derivations \Cref{sec:theory_prediction} for the NBP. We refer the interested reader to \citet{camerlenghi2022scaled,masoero2022more} for details on the estimators. We use code available at \url{https://github.com/lorenzomasoero/ScaledProcesses/} to fit these models and produce the corresponding estimates.

\subsection{Beta-binomial (BB) and beta-geometric (BG) predictors}

The beta-binomial and beta-geometric models are (finite dimensional) Bayesian generative models for trigger data which work by imposing a pre-determined upper bound $\unittotal_{\infty}$ on the number of units in the population, and assuming that for every unit $\unitidx=1,\ldots,\unittotal_{\infty}$ there exists a corresponding rate $\theta_n$ which governs the unit activity. In particular, these models assume
\[
    \theta_\unitidx \sim \mathrm{Beta}(\alpha,\beta).
\]
The beta-binomial model then assumes that we can observe for every day $\dayidx$ of experimentation and every unit $\unitidx$ where the unit triggered in the experiment on that day, and postulates
\[
    \triggerdata{\dayidx, \unitidx} \mid \theta_1,\ldots,\theta_{\unittotal_{\infty}} \sim \mathrm{Bernoulli}(\theta_\unitidx),
\]
i.i.d.\ across days $\dayidx$ for the same unit $\unitidx$, and independently across different units. The beta-geometric model instead assumes that we can only observe for every unit $\unitidx$ the ``first trigger date'' $\dayidx_\unitidx$, and postulates
\[
    \triggerdata{\unitidx} \mid \theta_1,\ldots,\theta_{\unittotal_{\infty}} \sim \mathrm{Geometric}(\theta_\unitidx).
\]
For the BB model, we use the estimator provided in \citet[Section 1]{ionita2009estimating} to produce $\prednews{\pilotdays}{\followupdays}$. For the BG model, we adopt the Monte Carlo approach devised in \citet[Section 3]{richardson2022bayesian} to produce the corresponding estimates. We provide code to fit these models and produce the corresponding estimates.

\subsection{Jackknife (J) predictors}

Jackknife estimators have a long history in the statistics literature, dating back to \citet{quenouille1956notes,tukey1958bias}. Here we consider the jackknife estimators developed by \citet{gravel2014predicting} who extended the work of \citet{burnham1979robust}. In particular, the $k$-th order jackknife is obtained by considering the first $k$ values of the resampling frequency spectrum; that is, by adequately weighting the number of users who appeared exactly $1, 2, \ldots, k$ times in the experiment. We adapt the code provided in \url{https://github.com/sgravel/capture_recapture/tree/master/software} to form our predictions.

\subsection{Good-Toulmin (GT) predictors}

Good-Toulmin estimators date back to the seminal work of \citep{good1956number}. Here, we adapt the recent approach of \citet{chakraborty2019using} for the problem of predicting the number of new genetic variants to be observed in future samples to online randomized experiments (in particular, we use the estimators provided in Equation (6) of the supplementary material). To form these predictions, we adapt to our setting the code provided in \url{https://github.com/c7rishi/variantprobs}.

\subsection{Linear programming predictors}

Linear programs have been used for rare event occurrence ever since the seminal work of \citet{efron1976estimating}. Here, we adapt to our setting the predictors proposed in \citet{zou2016quantifying} via the \texttt{UnseenEST} algorithm. We adapt the implementation provided by the authors at \url{https://github.com/jameszou/unseenest} to perform our experiments.
\section{Additional experiments} \label{sec:app_exp}

In this section, we provide additional experimental results.

\subsection{Synthetic data}

\subsubsection{Data from the model}

We here provide additional visualizations and results using data drawn from the model, adopting the marginal generative scheme of \Cref{sec:theory_sampling}. Python code to replicate and extend our analysis is provided.

\begin{figure}[ht]
    \centering
    \includegraphics[width=\linewidth]{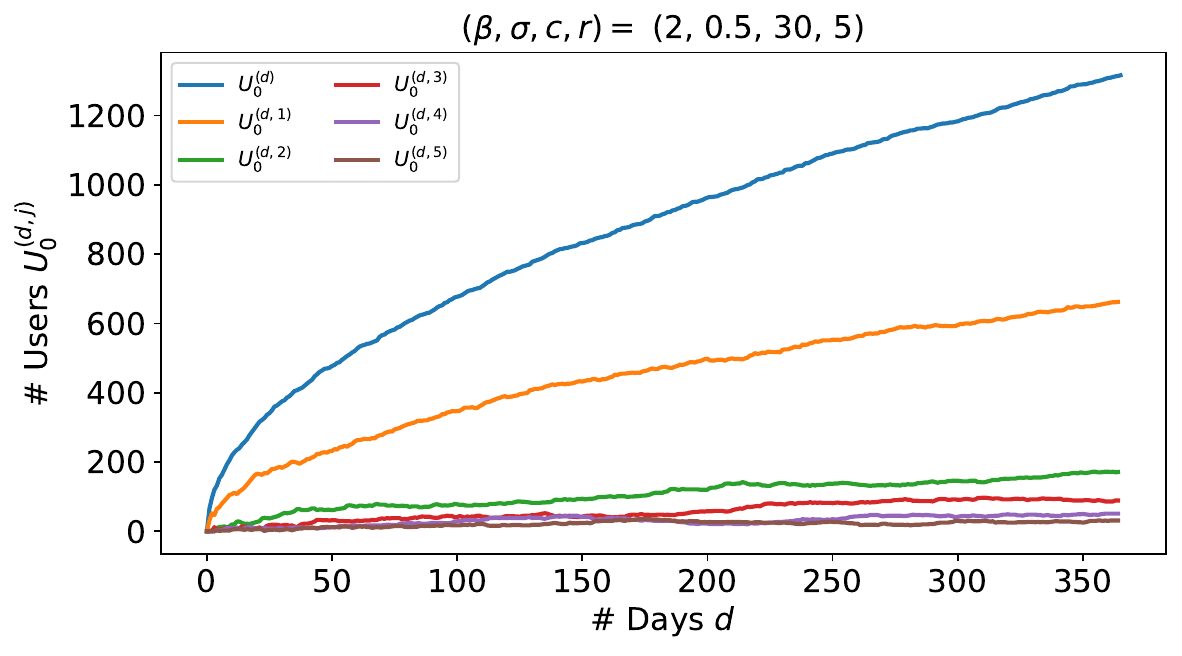}
    \caption{Accumulation curves for the number of new users for data drawn from the model (\Cref{eq:model}). Here, $\mass, \tail, \tilting, r = (2, 0.5, 30, 5)$ and $D=365$.}
    \label{fig:app_synthetic_accumulation}
\end{figure}

We here fix the model's hyperparameters to be $\mass, \tail, \tilting, r = (2, 0.5, 30, 5)$. We first show a representative draw from the model using this configuration and letting $D=365$ --- a hypothetical year of sampling. The corresponding sample statistics $\news{0}{\dayidx, \freq}$ are reported in \Cref{fig:app_synthetic_accumulation}.

\begin{figure}[ht]
    \centering
    \includegraphics[width=\linewidth]{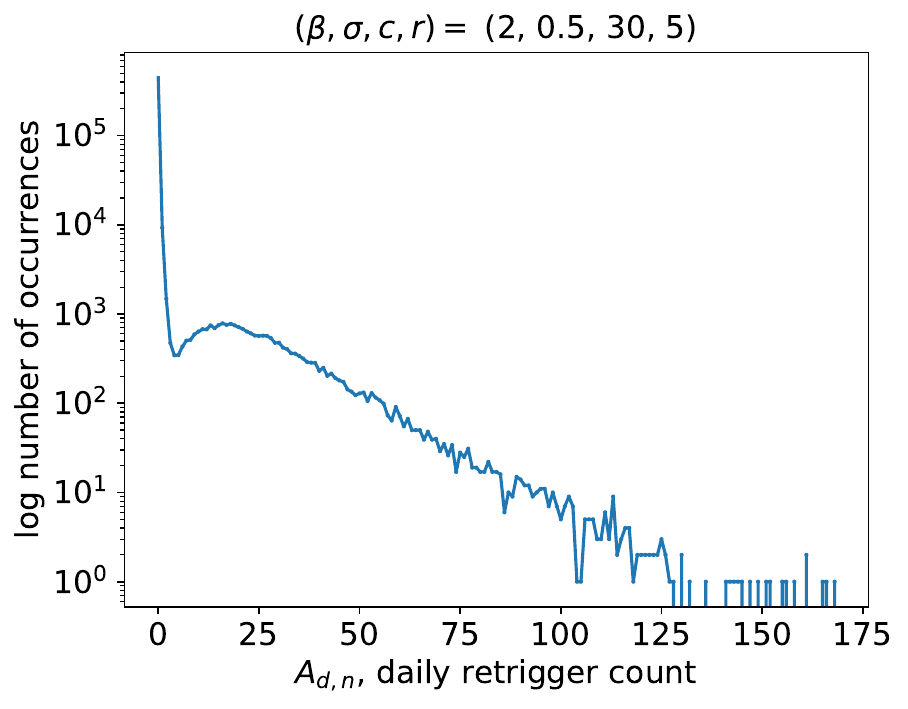}
    \caption{Spectrum of daily re-occurrences for the data already considered in \Cref{fig:app_synthetic_accumulation}.}
    \label{fig:app_synthetic_spectrum}
\end{figure}

We plot in \Cref{fig:app_synthetic_spectrum} the spectrum of daily re-occurrences. Specifically, we plot for each value $a = 1,2,
\dots$ (horizontal axis) the number of day-user pairs $(\dayidx, \unitidx)$ in the realized matrix $A$ of trigger occurrences that has value equal to $a$ (vertical axis, log scale).

\begin{figure}[ht]
    \centering
    \includegraphics[width=\linewidth]{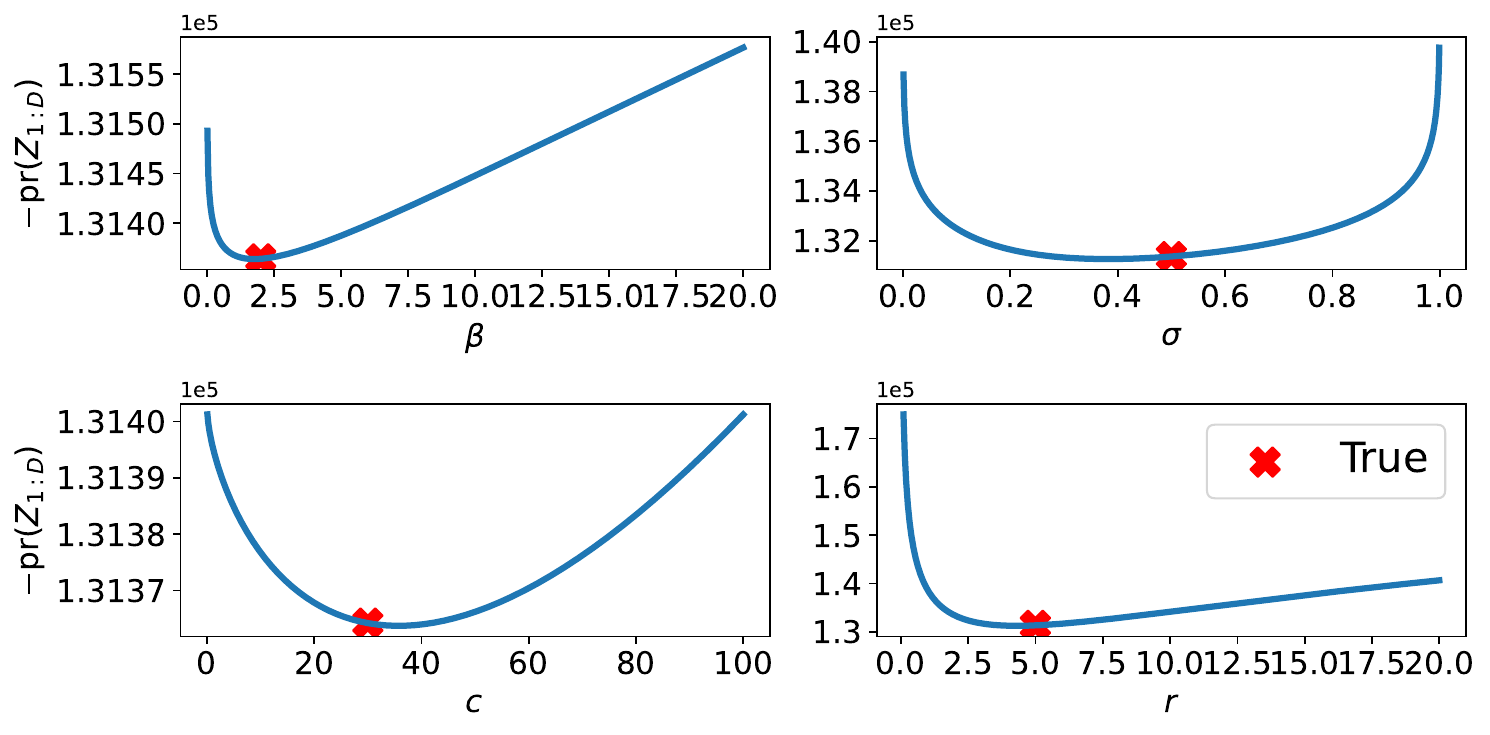}
    \caption{Evaluation of the log-likelihood in neighbourhoods of the true values for the same data already considered in \Cref{fig:app_synthetic_accumulation}.}
    \label{fig:app_synthetic_log_like}
\end{figure}

Next, we plot in \Cref{fig:app_synthetic_log_like} the profile of the negative (log) likelihood function around the true value of the parameters. Specifically, in each subplot, we evaluate the negative log-likelihood in the right and left neighbourhoods of the true value, shifting one coordinate at the time. We see that for all parameters, the negative log-likelihood is at least locally convex. We adopt a derivative-free optimization routine, which empirically robustly finds good values of the hyperparameters for the prediction task at hand. 

\begin{figure}[ht]
    \centering
    \includegraphics[width=\linewidth]{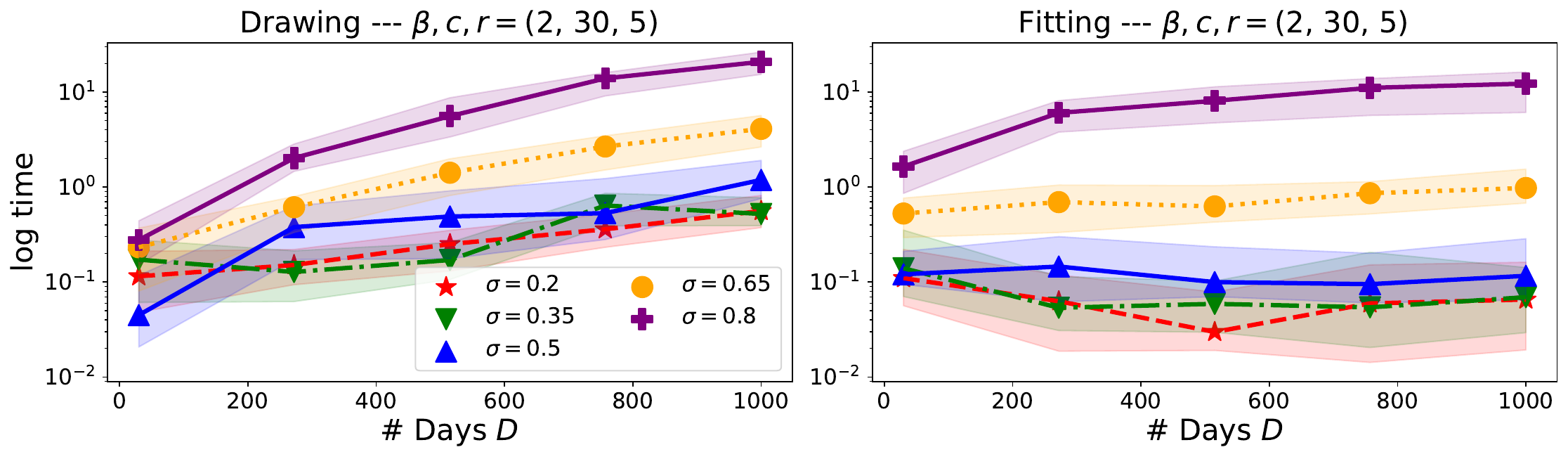}
    \caption{Drawing (left) and fitting (right) runtime for the model of \Cref{eq:model}, as the duration $D$ increases, across different choices of the hyperparameters.}
    \label{fig:app_synthetic_runtime_sigma}
\end{figure}

\begin{figure}[ht]
    \centering
    \includegraphics[width=\linewidth]{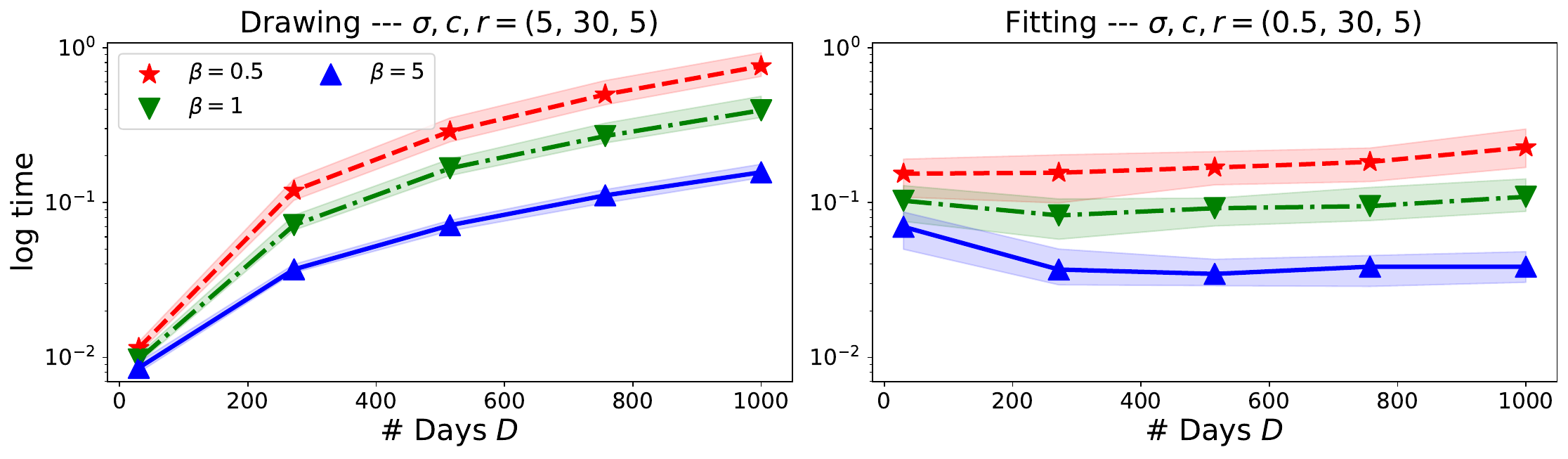}
    \caption{Drawing (left) and fitting (right) runtime for the model of \Cref{eq:model}, as the duration $D$ increases, across different choices of the hyperparameters. }
    \label{fig:app_synthetic_runtime_beta}
\end{figure}

Last, we provide run time information. Specifically, we show how drawing and fitting times scale as a function of the number of days $D$ increases and for different configuration of the hyperparameters. In \Cref{fig:app_synthetic_runtime_sigma} we consider different values of the tail parameter $\tail$. Higher values of the parameter induce a larger number of unique users, and result in larger data, translating in longer draw and fitting times. Lower values of the mass parameter $\mass$ have a similar effect, as displayed in \Cref{fig:app_synthetic_runtime_beta}. Generally speaking, fitting the hyperparameters is computationally cheap, since it only amounts to either a regression problem on four parameters, or a likelihood maximization with respect to four parameters, where the likelihood function is available in closed form and can be efficiently evaluated. For each configuration of the hyperparameters and the duration $D$ considered in \Cref{fig:app_synthetic_runtime_sigma,fig:app_synthetic_runtime_beta} we repeat the experiment $M=30$ times and report the median elapsed time together with a centred interval of elapsed times of width 80\%, to assess the variability in our estimates.

\subsection{Real data}

We provide some additional experiments on real data. Here we consider the same proprietary dataset displayed in \Cref{fig:accuracy_1}, but for a subset of 1{,}340 comparisons for which we can extrapolate up to $\followupdays=35$. We retain again $\pilotdays=7$ days of triggering data for training. Even for this longer horizon period, the BNP methods perform best.

\begin{figure}[ht]
    \centering
    \includegraphics[width=\linewidth]{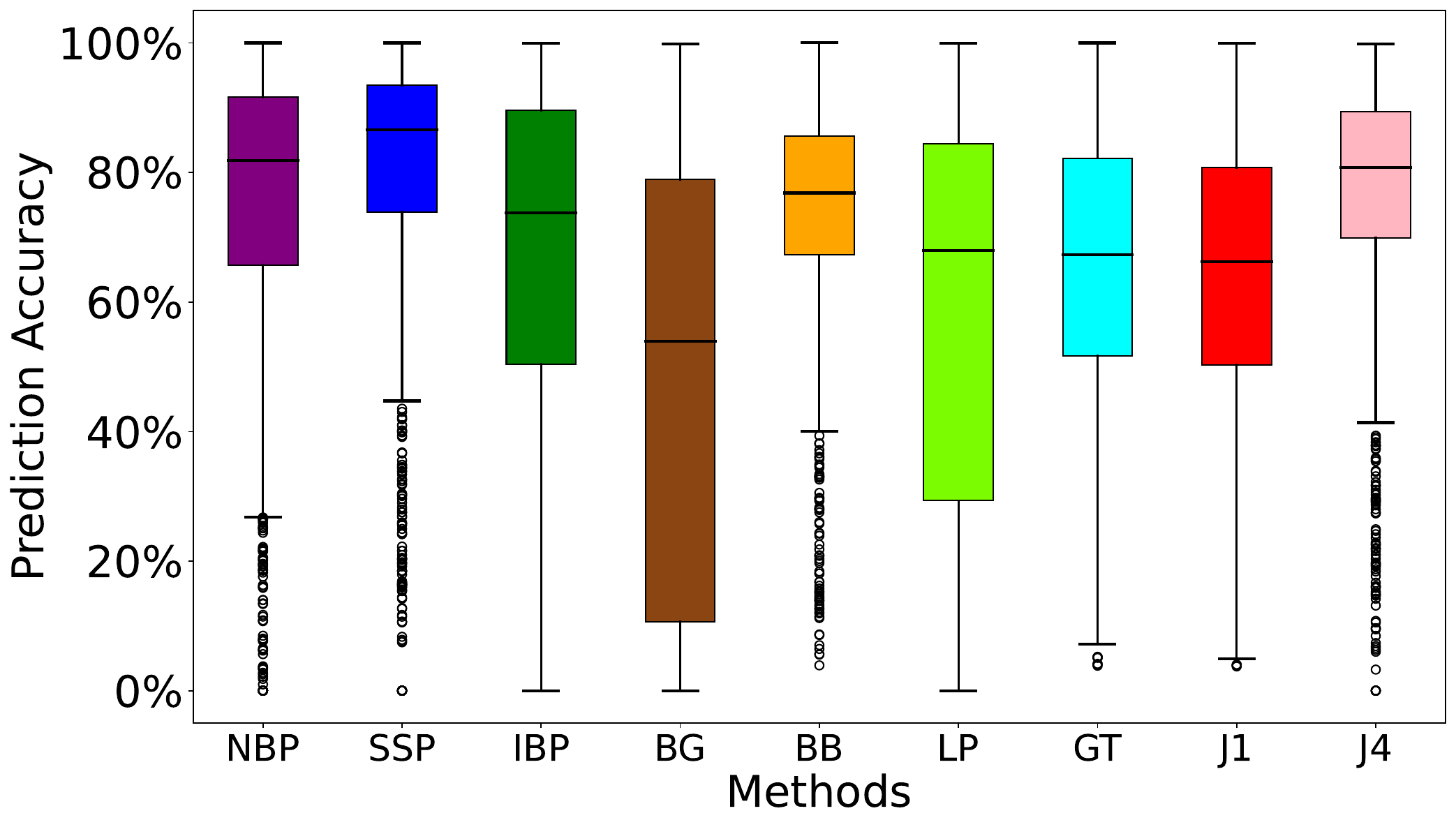}
    \caption{Prediction accuracy $v_{7}^{(35)}$ on proprietary data.}
    \label{fig:app_real_42}
\end{figure}
\end{document}